\documentclass[aps,prx,reprint,showpacs,superscriptaddress,floatfix,pra]{revtex4-1}

\usepackage[english]{babel}
\usepackage{comment}
\usepackage{graphics,amssymb,amsmath,epsfig,color,textgreek}
\usepackage{graphicx}
\usepackage[dvipsnames]{xcolor}
\usepackage{dcolumn}% Align table columns on decimal point
\usepackage{bm}% bold math
\usepackage{hyperref}
\hypersetup{colorlinks,linkcolor=blue,anchorcolor=blue,citecolor=blue,urlcolor=blue}
\usepackage{braket}
\usepackage[normalem]{ulem}
\usepackage{url}
\usepackage{cancel}
\usepackage{diagbox}
\usepackage{lipsum}
\usepackage{cleveref}
\usepackage{subfigure}
\usepackage{tikz}
\usepackage{orcidlink}
\usetikzlibrary{arrows.meta,positioning,calc,fit,backgrounds,decorations.pathmorphing}

\newcommand{\nocontentsline}[3]{}
\let\origcontentsline\addcontentsline
\newcommand\stoptoc{\let\addcontentsline\nocontentsline}
\newcommand\resumetoc{\let\addcontentsline\origcontentsline}

\setcitestyle{numbers,square,comma,sort&compress}

\DeclareUnicodeCharacter{2009}{\,} 
\newcommand{\dd}{\mathrm{d}}
\newcommand{\Cbb}{\mathbb{C}}
\newcommand{\Rbb}{\mathbb{R}}
\newcommand{\Tcal}{\mathcal{T}}
\newcommand{\Ocal}{\mathcal{O}}

\newcommand{\Z}{\mathbb{Z}}

\crefformat{figure}{Fig.~#2#1#3}
\crefformat{table}{Tab.~#2#1#3}
\crefformat{equation}{Eq.~#2#1#3}
\crefformat{appendix}{App.~#2#1#3}

\begin{document}

\title{Hybrid continuous-discrete-variable quantum computing: a guide to utility}

\author{A.F. Kemper\,\orcidlink{0000-0002-5426-5181}}
\email{akemper@ncsu.edu}
\affiliation{Department of Physics, North Carolina State University, Raleigh, North Carolina 27695, USA}

\author{Antonios Alvertis}
\affiliation{KBR, Inc., NASA Ames Research Center, Moffett Field, California 94035, United States
}

\author{Muhammad Asaduzzaman}
\affiliation{Department of Physics, North Carolina State University, Raleigh, North Carolina 27695, USA}

\author{Bojko N. Bakalov}
\affiliation{Department of Mathematics, North Carolina State University, Raleigh, North Carolina 27695, USA}

\author{Dror Baron}
\affiliation{Department of Electrical and Computer Engineering, North Carolina State University, Raleigh, North Carolina 27695, USA}

\author{Joel Bierman}
\affiliation{Department of Electrical and Computer Engineering, North Carolina State University, Raleigh, North Carolina 27695, USA}

\author{Blake Burgstahler}
\affiliation{Department of Computer Science, North Carolina State University, Raleigh, North Carolina 27695, USA}

\author{Srikar Chundury}
\affiliation{Department of Computer Science, North Carolina State University, Raleigh, North Carolina 27695, USA}

\author{Elin Ranjan Das}
\affiliation{Department of Electrical and Computer Engineering, North Carolina State University, Raleigh, North Carolina 27695, USA}
\affiliation{Pacific Northwest National Laboratory, Richland, WA, USA, 99354}

\author{Jim Furches}
\affiliation{Pacific Northwest National Laboratory, Richland, WA, USA, 99354}

\author{Fucheng Guo}
\affiliation{Department of Computer Science, North Carolina State University, Raleigh, North Carolina 27695, USA}

\author{Raghav G. Jha}
\affiliation{Department of Physics, North Carolina State University, Raleigh, North Carolina 27695, USA}

\author{Katherine Klymko}
\affiliation{NERSC, Lawrence Berkeley National Laboratory, Berkeley, California 94720, USA}

\author{Arvin Kushwaha}
\affiliation{Department of Physics, North Carolina State University, Raleigh, North Carolina 27695, USA}
\affiliation{Department of Mathematics, North Carolina State University, Raleigh, North Carolina 27695, USA}

\author{Ang Li}
\affiliation{Pacific Northwest National Laboratory, Richland, WA, USA, 99354}
\affiliation{Department of Electrical and Computer Engineering, University of Washington, Seattle, WA, USA, 98195}

\author{Aishwarya Majumdar}
\affiliation{Department of Electrical and Computer Engineering, North Carolina State University, Raleigh, North Carolina 27695, USA}

\author{Carlos Ortiz Marrero}
\affiliation{Pacific Northwest National Laboratory, Richland, WA, USA, 99354}
\affiliation{Department of Computer Science, Colorado State University, Fort Collins, Colorado 80524, USA}

\author{Shubdeep Mohapatra}
\affiliation{Department of Electrical and Computer Engineering, North Carolina State University, Raleigh, North Carolina 27695, USA}

\author{Christopher Mori}
\affiliation{Department of Electrical and Computer Engineering, North Carolina State University, Raleigh, North Carolina 27695, USA}

\author{Frank Mueller}
\affiliation{Department of Computer Science, North Carolina State University, Raleigh, North Carolina 27695, USA}

\author{Doru Thom Popovici}
\affiliation{Lawrence Berkeley National Laboratory, Berkeley, California 94720, USA}

\author{Tim Stavenger}
\affiliation{Pacific Northwest National Laboratory, Richland, WA, USA, 99354}

\author{Mastawal Tirfe}
\affiliation{Department of Physics, North Carolina State University, Raleigh, North Carolina 27695, USA}
\affiliation{Department of Mathematics, North Carolina State University, Raleigh, North Carolina 27695, USA}

\author{Norm M. Tubman}
\affiliation{NASA Ames Research Center, Moffett Field, California 94035, USA}

\author{Muqing Zheng}
\affiliation{Pacific Northwest National Laboratory, Richland, WA, USA, 99354}

\author{Huiyang Zhou}
\affiliation{Department of Electrical and Computer Engineering, North Carolina State University, Raleigh, North Carolina 27695, USA}

\author{Yuan Liu\,\orcidlink{0000-0003-1468-942X}}
\email{q\_yuanliu@ncsu.edu}
\affiliation{Department of Electrical and Computer Engineering, North Carolina State University, Raleigh, North Carolina 27695, USA}
\affiliation{Department of Computer Science, North Carolina State University, Raleigh, North Carolina 27695, USA}
\affiliation{Department of Physics, North Carolina State University, Raleigh, North Carolina 27695, USA}

\date{\today}

\begin{abstract}
Quantum computing has traditionally centered around the discrete variable paradigm.  A new direction is the inclusion of continuous variable modes and the consideration of a hybrid continuous-discrete approach to quantum computing.  In this paper, we discuss some of the advantages of this modality, and lay out a number of potential applications that can make use of it; these include applications from physics, chemistry, and computer science. We also briefly overview some of the algorithmic and software considerations for this new paradigm.
\end{abstract}

\maketitle

\tableofcontents

\section{Introduction}

\begin{figure*}
    \centering
    \includegraphics[width=0.95\linewidth]{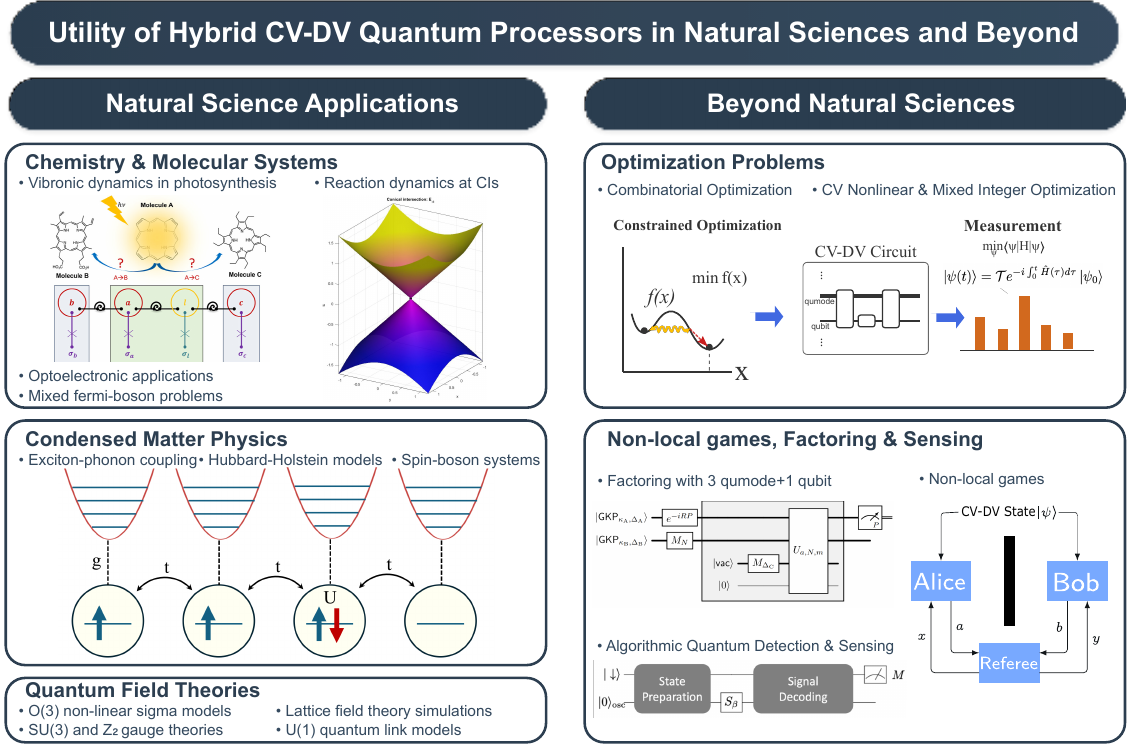}
    \caption{Applications of hybrid CV-DV quantum computing in Natural sciences and beyond. Part of the chemistry panel is adapted with permission from \cite{vu2025computational}. Copyright@2025 American Chemical Society.}
    \label{fig:application}
\end{figure*}

The last decade has seen a rapid rise in the availability, capabilities, and algorithms designed for quantum computers.  The move from the laboratory into an industry setting has further accelerated the already growing field, and has taken the theoretical developments over the past decades and put them to use in applications from stock portfolio optimization \cite{buonaiuto2023best} to simulations of quantum chromodynamics \cite{bauer2023quantum}.

The primary building block for these advances has been the \textit{qubit} --- a well-controlled two-level system, of any origin, that can be used to implement quantum algorithms and simulate other quantum systems. Since any two level system can be used as such, the options are varied and they include nuclear \cite{vandersypen2005nmr,jones2024controllingNMR} and electron spins \cite{dijkema2023twoqubit}, pairs of levels in natural atoms \cite{henriet2020quantum}, ions \cite{HAFFNER2008155,Bruzewicz_2019,Quantinuum_moses2023race,Katz_2023_prxq}, and color centers \cite{ColorCenters10.1063/5.0007444}, as well as synthetic atoms such as quantum dots \cite{RevModPhys.95.025003} and superconducting qubits \cite{Reagor2013,reagor_2016,PhysRevApplied.13.034032,
Rosenblum_TensofMilliseconds_PRXQuantum.4.030336,YYGao_QIP_bosonic_cQED,BlaiscQEDReviewRMP2020,Blais2020}, and various possible hybrid systems \cite{RevModPhys.85.623,andersen2015hybrid}.  These typically have an algebra similar to spins; some local (anti-)commutation relations but nothing broader.  Notably different are bosons and fermions, and some recent extensions have been proposed that build two-level systems out of these\cite{Braunstein_2005,Weedbrook2012,andersen2015hybrid,YYGao_QIP_bosonic_cQED}. These options have been explored to some extent for quantum simulations, but remain relatively unexplored in the context of quantum computation \cite{Bravyi_2002, obrien2018majorana, LloydQuantumComputation1999}.

Bosons are in particular different; a single bosonic mode or harmonic oscillator has a countable infinity of states and can therefore be characterized by a continuous variable (CV), in stark contrast to both qubits and fermions.  A recent development has been to hybridize the CV architecture with the traditional discrete variable (DV, qubit) architecture. These situations have naturally existed in both ion-trap and superconducting circuit systems, but have not yet been exploited.
Yet they 
deserve to be better known, because they offer a number of powerful advantages over more traditional qubit-only systems for quantum error correction \cite{albert_performance_2018,grimsmo_quantum_2020,Girvin_LesHouches_QEC}, quantum simulations \cite{Wang2020FCFs,WangConicalIntersection}, efficient state tomography \cite{feldman2022selective,Gullans_CV_Classicalshadows}, and potentially for quantum algorithms \cite{pati2000quantum,lomonaco2002continuous,pati2003deutsch,aaronson2011computational,lund2014boson}. These advantages are starting to be experimentally realized, and system development for such hybrid hardware is in the early stages of scaling up to large sizes. The superposition principle and the Born Rule of quantum mechanics ensure that DV systems have both digital and analog characteristics. These characteristics of hybrid CV-DV systems bring unique opportunities for demonstrating quantum utility, not only against classical computers but also against qubit-based quantum computers.

With the field heading towards a broader incorporation of mixed CV-DV paradigm, it is incumbent upon us to consider the possible advantages that can be gained by current and future developments in this direction.  The purpose of this work is to outline several problems from natural and computational science that could benefit from being treated in a mixed-mode representation, rather than purely one or the other.  We will present examples from physics and chemistry, where it is quite natural to find fermions (typically electrons) coupled to bosonic modes such as lattice vibrations (phonons), coherent light fields (photons) or other emergent bosonic modes. Beyond the physical sciences, we will consider example problems from optimization, factoring, data compression, differential equations, and sensing. For each of these, we will discuss the problem, which may appear in one or more guises, and outline the problem in the context of CV/DV hardware. We finish with a discussion on fault-tolerance and the current state of algorithms and software for CV-DV systems.

We begin with a discussion of the general advantages of hybrid CV-DV systems, and how problems of interest may be mapped on such hardware in Sec. \ref{sec:advantage-mapping}. Following that, we highlight a number of open problems of interest from the natural science in Sec. \ref{sec:natural-science} to general computing problems in Sec. \ref{sec:other-computing} that could make use of a hybrid CV-DV architecture, outlining the problem structure, desired observables, and possible necessary algorithms. We briefly suggest novel algorithms, software, and techniques that are enabled by a hybrid architecture in Sec. \ref{sec:alg-soft-tech} before making some closing remarks in Sec. \ref{sec:conclusion}.

\section{General Advantages of Hybrid CV-DV Computation}
\label{sec:advantage-mapping}

The advantages of hybrid CV-DV quantum computation for general computational problems are multifold. The first comes from the continuous or infinite-dimensional nature of the state space of a qumode. This feature allows for mapping of any continuous degrees of freedom (or state variables) in the underlying problems directly to quadratures of one or a few qumodes, as opposed to many qubits. This can significantly reduce mapping overhead without unnecessary discretization errors. As is highlighted in Secs. III and IV, examples of this kind of advantage include mixed fermi-boson matter and quadratic unconstrained binary optimization.

The second advantage comes from the unique quantum statistics of the CV hardware, which makes simple CV operations extremely non-trivial to compute in the DV computational basis. For example, bosonic creation and annihilation operations on Fock states can be viewed as performing a square-root arithmetic operation on all positive integers. Moreover, free evolution of CV systems, such as bosonic oscillators or quantum rotors, can be viewed as performing the quantum Fourier transform, a key computational primitive for quantum computation \cite{Liu_2025}. This advantage directly facilitates the simulation of boson-containing quantum matter~\cite{liu2024hybrid,crane2024hybridoscillatorqubitquantumprocessors,huh_boson_2015,sparrow_simulating_2018,HU2018293,Wang2020FCFs,WangConicalIntersection}, where mapping such operations into DV frameworks is otherwise prohibitively costly. 

Another advantage of the hybrid approach lies in the choice of the universal gate set. 
As is well known, due to the Gottesman--Knill theorem, the set of Clifford gates 
is not universal and must be supplemented by the $T$-gate to achieve universal quantum computing. 
Likewise in the CV case, one cannot rely solely on Gaussian gates; 
one must either employ the cubic (or higher-order) phase gate or use photon-number detection 
as the non-Gaussian element (similar to the Knill--Laflamme--Milburn criterion). In the hybrid CV-DV setup, we can achieve universal quantum computing without using either non-Gaussian CV or non-Clifford DV gates just by having access to controlled position displacement in the $Z$- and $X$-basis, i.e., $U(\alpha) = 
    \exp(-i\alpha\, \sigma_z \otimes X)$, $U(\beta) = \exp(-i\beta\, \sigma_x \otimes X)$,  
    and controlled momentum displacement in the $Z$-basis, i.e., $U(\gamma) = \exp(-i\gamma\, \sigma_z \otimes P)$.  Together with the ability for two qubits to couple to a common oscillator mode, and for a single qubit to couple to multiple oscillators, these operations suffice for universality in the hybrid setting~\cite{Lloyd1999hybrid}. Depending on the physical platform, this route may be experimentally more accessible than other known universal gate constructions, providing a potential implementation advantage.

Last but not least, from a fault-tolerant perspective, quantum information can be encoded redundantly in hybrid CV-DV devices composed of multiple qubits and qumodes. This redundancy in a high-dimensional space with a relatively simple hardware error mechanism allows the encoding and manipulation of logical DV \cite{Joshi_2021,cai_bosonic_2021,GrimsmoPuri_GKP_2021,DissipativeCatproposal,albert_performance_2018,grimsmo_quantum_2020,BinomialCodes,Royer_2022} and CV \cite{noh2020encoding,wu2021continuous,Wu_2023,Hanggli_2022} quantum information with relatively low overhead. Break-even for logical qubits in these bosonic QEC codes has recently been experimentally demonstrated~\cite{Ofek2016,ni2022beating, Sivak_GKP_2022,lachance2024autonomous}. The encoding of a logical subspace in multiple CV systems that \emph{preserves the unique CV statistics (bosonic or rotor-like)} is crucial for hybrid CV-DV quantum advantages over purely DV systems, because it allows long coherent quantum computation to be performed on virtual devices with CV characteristics.

\subsection*{Mapping strategies}

The advantages of hybrid CV-DV computation rely on mapping strategies from the problem's degrees of freedom to logical hybrid CV-DV degrees of freedom. There are two steps to consider in the mapping: i) pick state variables of the underlying problem to represent the problem; ii) choose a computational basis on the hybrid CV-DV hardware to encode the given state variables. Step i) has also been considered for quantum computation with DV devices, although often one should consider both i) and ii) holistically for the best mapping strategy. The guiding principle for the best mapping strategy is one such that the required computational process can be most succinctly \emph{represented} (relevant to space resource cost) on and \emph{performed} (relevant to circuit depth and time resource cost) by the hybrid CV-DV quantum hardware.

For qubits, the choice of computational basis is relatively boring, because most likely one chooses the Pauli-Z eigenstates $\{ \ket{0}, \ket{1} \}$. An alternative computational basis such as $\{ \ket{+}, \ket{-} \}$ can be used, but this does not often lead to significant computational advantages compared to the Z-basis. % \textcolor{red}{YL needs to discuss entangled basis simulation on qubits, and how that relates to CV basis.} 
What makes CV quantum hardware interesting is that there are many different sets of bases to serve as computational bases where the complexity of solving the same problem can be very different depending on the mapping.

The states of a bosonic mode's Hamiltonian $\hat{H} = \hbar \omega \hat{n}$ define a countably infinite set of basis vectors $\{ \ket{n} \}$ for $(n = 0, 1, 2, \cdots)$, called the number states or Fock states, where $\hat{n} = b^\dagger b$ is the photon number operator with $n$ the associated eigenvalue (we use $\hbar \omega = 1$ in the rest of the text for notation simplicity) describing how many quanta of excitations are populated, and $b^\dagger, b$ are the creation and annihilation operators that add or remove one quantum of excitation into or from the mode. These number vectors $\{ \ket{n} \}$ can be used to encode any countably infinite (or a large number of finite) discrete degrees of freedom (DOFs) in any underlying problem. For the case of finite DOFs, one can simply truncate the Fock basis to a finite value $N_{\rm max}$ and only use the states $\ket{n}$ with $n \le N_{\rm max}$ for the computation. This is similar to using an oscillator as a qudit. The practical implementation of such a computation within the low-energy subspace requires some care \cite{QuditsfromOscillatorsPhysRevA.104.032605}, but it is still doable.

In addition to the countably infinite number of Fock states, the position or momentum states $\ket{x}$ or $\ket{p}$ ($x, p \in \mathbb{R})$ can be directly used as a computational basis to represent fully continuous DOFs of the problem, where $\ket{x}$ or $\ket{p}$ are defined as the eigenstates of the position operator $\hat{x}$ or the momentum operator $\hat{p}$ with eigenvalue $x$ or $p$, i.e., $\hat{x} \ket{x} = x \ket{x}$ and $\hat{p} \ket{p} = p \ket{p}$. For example, many algorithms involve integration over a continuous variable, which can be mapped natively to the oscillator's position or momentum variable \cite{bell2025codesigningeigensingularvaluetransformation}.

CV systems also enable unconventional computational bases that are non-orthogonal and over-complete such as the set of all coherent states $ \ket{\alpha}$ for $\alpha \in \mathbb{C}$. The simplicity of these coherent state bases is that the action of Gaussian operations (unitaries generated by linear and quadratic powers of $\hat{x}$ and $\hat{p}$) on $\ket{\alpha}$ is very easy to follow. This makes the matrix elements of Gaussian operators simple. However, the difficulty lies in their non-orthogonality, which means that one needs to keep track of the overlap integrals between all coherent states during the computation.

\section{Natural Science Applications}
\label{sec:natural-science}

Simulating physical systems has been one of the main utilities of quantum computers to achieve a practical quantum advantage \cite{quantum_initiative_US,quantum_flagship_Europe, Bauer_SimulationHEP_2022, Cao2019, RevModPhys.86.153}, as originally proposed by Feynman \cite{feynman1985quantum}. In chemistry and condensed matter physics, a wide variety of systems can be generically described as a system of interacting fermions coupled to some set of bosonic modes---these range from simple phonon modes that can be described by a Holstein-type coupling to complex magnetic excitations and molecular vibrations.  These modes can be treated within a discrete or a continuous framework, depending on the particular approach. Although classical hybrid Monte Carlo simulations have shown promise in studying these systems, they retain the usual problems with Monte Carlo (e.g., sign problems). Explicit examples of problems within this class include the interaction between strongly coupled phonons and superconductivity in the high-Tc cuprates \cite{varma2020colloquium}, fermionic systems coupled to the U(1) gauge field of electromagnetism (light-matter interaction), or the SU(3) field in quantum chromodynamics \cite{bauer2023quantum}. Although these systems and physics have been studied extensively for decades, questions remain unanswered due to the fundamental limitations of classical simulations on both the fermionic and bosonic sides.  And, while DV qubits could potentially overcome problems with fermions, they are not well suited to treating bosonic fields due to the difficulty in mapping bosonic systems to qubits \cite{shaw2020quantum,sawaya2019quantum} as well as the associated challenges in implementing the action of bosonic field operators (for example the creation and annihilation operators) \cite{jordan2012quantum,jordan2014quantum} on qubit-based quantum computers \cite{bhaskar2015quantum,soeken2017hierarchical}.

In contrast to the situation with purely DV modes, where the difficulty of simulating bosonic matter arises due to a very large number of quantum gates needed to accurately synthesize the square root factors~\cite{soeken2017hierarchical,bhaskar2016quantum} from the bosonic field operators, the field operators are natively available in hardware containing bosonic modes. Therefore, fermi-boson mixtures can be efficiently mapped to hybrid CV-DV quantum processors. The efficient mapping of bosonic modes has recently been experimentally demonstrated by a construction of a highly hardware-efficient quantum simulation of the Franck--Condon vibrational spectra of small molecules using Gaussian boson sampling with optical modes \cite{huh_boson_2015,sparrow_simulating_2018} and with microwave modes \cite{HU2018293,Wang2020FCFs} representing the mechanical vibrational modes. These simple devices with natively available bosonic modes accurately performed simulations that would be impossible on any currently existing superconducting qubit-only quantum computer \cite{Wang2020FCFs}.  A related experiment has recently simulated dissipative molecular quantum dynamics near a conical intersection \cite{WangConicalIntersection}. These recent 
advances demonstrate the great potential of hybrid CV-DV quantum processors for quantum simulations.

\subsection{Mixed fermi-boson problems from chemistry and physics}

\begin{figure}[h]
    \centering
    \includegraphics[width=0.49\textwidth]{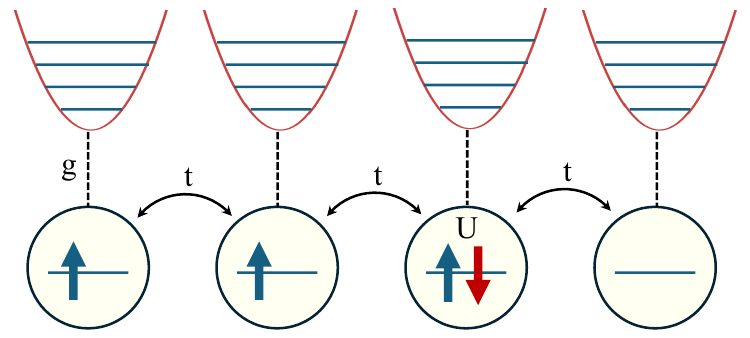}
    \caption{A prototypical mixed-fermi boson model: the Hubbard-Holstein model.  Electrons (arrows) are coupled to local Einstein oscillators, can hop between sites, and are subject to an on-site Coulomb repulsion.  Note that the dimension of the Hilbert space associated to each site is 4.}
    \label{fig:hubbard_holstein}
\end{figure}

Many problems involving a mixture of fermions and bosons share a common Hamiltonian: individual terms for the discrete (fermion) and continuous (boson) parts, and an interaction term.  As a prototypical example, we may consider a generalized Tavis-Cummings model,
\begin{align}
    \mathcal{H}_{\mathrm{TC}} &= 
        \underbrace{\sum_{i} \varepsilon c^\dagger_i c_i + \sum_{i,j} h_{ij} c^\dagger_i c_j}_{\text{discrete}}  \nonumber \\
        &+ \sum_k \underbrace{\omega_k a_k^\dagger a_k}_{\text{continuous}}  + 
        \underbrace{g \sum_{i,k} \left( c^\dagger_i a_k + c_i a_k^\dagger \right)}_{\text{interaction}}.
        \label{eq:fermi-boson-hamiltonian}
\end{align}
Here we have in mind a system of electrons with associated creation (annihilation) operators $c_i^\dagger$ ($c_i$) expressed
in some natural basis indexed by $i$ --- this may represent, for example, orbitals or lattice sites in a model motivated by physics or chemistry --- with a resulting on-site energy term $\varepsilon$ and a kinetic energy $h_{ij}$.  Similarly, the bosons are indexed by $k$ and have their own creation (annihilation) operators $a_k^\dagger$ ($a_k$).  The form of the interaction term
varies from problem to problem~\cite{hohenadler2004quantum,sengupta2003peierls,kim2024semiclassical};
the one we have represented here is reminiscent of an all-qubit-to-all qumode coupled Tavis-Cummings model.
Other types of coupling may include, for example, a Holstein-type coupling $g c^\dagger_i c_i (a_k+a_k^\dagger)$ as shown in Fig.~\ref{fig:hubbard_holstein}; the particular problem under consideration will dictate the form. The discrete variable portion of the Hamiltonian may also have non-fermionic character as long as the Hilbert space is limited to two levels, such as in spin-boson models or in the exciton-phonon models discussed below.

\subsection{Spin-boson models}
The general spin–boson Hamiltonian \cite{leggett1987dynamics, weiss2012quantum} for a system comprising $N$ spins (two-level systems) interacting with a bosonic environment can be written as
\begin{equation}
H = H_S + H_B + H_{SB},
\end{equation}
where $H_S$ describes the system degrees of freedom, $H_B$ the bosonic bath, and $H_{SB}$ their mutual coupling.

The system Hamiltonian is
\begin{equation}
H_S = \sum_{i=1}^N \frac{\epsilon_i}{2}\sigma_i^z + \frac{\Delta_i}{2}\sigma_i^x
+\sum_{i<j} J_{ij}\sigma_i^z \sigma_j^z ,
\end{equation}
where $\epsilon_i$ is the energy bias of spin $i$, $\Delta_i$ is its tunneling amplitude, and $\sigma_i^{x,z}$ are Pauli operators acting on spin $i$. The coefficients $J_{ij}$ represent pairwise Ising-type couplings that allow for spin–spin correlations or collective dissipation effects.

The bath Hamiltonian represents a collection of independent harmonic oscillators,

\begin{equation}
H_B = \sum_k \omega_k b_k^\dagger b_k,
\end{equation}
where $b_k^\dagger$ ($b_k$) creates (annihilates) a boson of mode $k$ with frequency $\omega_k$. This formulation assumes a continuum of bath modes that can absorb or emit energy from the system.

The system–bath interaction Hamiltonian is typically taken to be linear in the bosonic coordinates,
\begin{equation}
H_{SB} = \sum_{i=1}^N \sum_k g_{ik} \sigma_i^z \left( b_k^\dagger + b_k \right),
\end{equation}
where $g_{ik}$ denotes the coupling strength between spin $i$ and bosonic mode $k$. This coupling form ($\sigma_i^z$) is the most common found in the literature and leads to decoherence primarily through dephasing (although other coupling channels are sometimes considered for relaxation processes).

The environmental influence is completely characterized by the spectral density,
\begin{equation}
J(\omega) = \pi \sum_{k} g_{k}^2  \delta(\omega - \omega_k),
\end{equation}
which encodes how strongly each bath frequency $\omega$ is coupled to the system. Depending on its functional form, different dynamical regimes arise, ranging from coherent oscillations to overdamped localization.

Spin–boson models constitute a central framework for describing dissipative quantum dynamics in two-state systems interacting with environments. The bosonic bath represents the environmental degrees of freedom as a continuum of harmonic modes coupled linearly to the system coordinates. This paradigm has been extensively used to study various physical processes, including the motion of light particles in metals and disordered solids \cite{de2017dynamics}, chemical reactions involving electron transfer processes \cite{grifoni1998driven}, decoherence and relaxation in superconducting qubits \cite{makhlin2001quantum, paladino20141}, and vibrational dynamics and exciton–phonon coupling in molecular aggregates \cite{kolli2012fundamental}.

Beyond its role as a prototypical open-quantum-system model, the spin–boson Hamiltonian also serves as a building block in quantum simulation~\cite{porras2004effective}, quantum thermodynamics~\cite{leggio2015quantum}, and quantum information processing~\cite{makhlin2001quantum, paladino20141}, where controlled environments (e.g., engineered phonon or photon baths) allow systematic exploration of non-Markovian~\cite{rivas2014quantum} and strong-coupling~\cite{ishizaki2009theoretical} effects.

\subsection{Modeling fermi-boson and spin-boson models}
Quantum simulators have realized simplified Holstein and spin--boson models using trapped ions and superconducting circuits \cite{huh2017vibronic, RevModPhys.86.153}, making use of existing quantum algorithms including Trotter--Suzuki decomposition, qubitization, and quantum trajectory methods \cite{lloyd1996universal,low2017optimal}.
However, treating these problems on purely DV platforms is difficult:
large active spaces require many qubits, while bosonic vibrational modes need high Fock cutoffs, scaling qubit demands rapidly. Non-Markovian baths with structured spectral densities are costly to encode digitally. Moreover, Lindblad evolution is not natively unitary, making open-system simulation resource intensive and requiring
the dynamics to be approximated by Stinespring dilation, collision models, or quantum trajectories 
 \cite{RevModPhys.92.015003,RevModPhys.89.015001}.
Hybrid architectures overcome these difficulties by assigning qubits to discrete electronic states and CV modes (oscillators, cavities, ion motion) to vibrations and baths. This naturally realizes the mixed fermi-boson or spin–boson  Hamiltonians without large truncations. One can engineer the CV modes to reproduce structured spectral densities, and Lindblad-type dissipation integrates seamlessly in this framework. This was shown recently in Ref.~\onlinecite{doi:10.1021/acs.jctc.5c00315}, where oscillator–qubit partitioning captured coherent vibronic coupling and dissipative nonadiabatic dynamics with far fewer resources than qubit-only approaches.

In general, there are two classes of observables.  The first is a single point measurement, i.e., observables of the form
$\mathrm{tr} \left[ \rho \mathcal{O} \right]$ for some operator $\mathcal{O}$ and a (potentially time-dependent) state $\rho$.  These include site populations, bosonic occupations, coherences, DC currents, etc.  The second class are two-time
correlation functions
\begin{align}
C(t) =   \mathrm{tr}\left[ \rho A(t) B(0) \right],
\end{align}
which illustrates the need to be able to have time-dependent correlation functions. 
Alternatively, we may express the entire problem in the Fourier
domain, where instead a function of the Hamiltonian needs to be evaluated
\begin{align}
    \mathrm{Re}\ \sigma(q,\omega) = \mathrm{tr}\left[ \rho A\frac{1}{\omega - \mathcal{H} + E_0 + i 0^+} B\right],
\end{align}
where $0^+$ is an infinitesimal regularization factor and $E_0 = \mathrm{tr} \left[ \rho H \right]$.
Key observables of this type include many experimentally accessible spectra, such as photoemission spectroscopy, optical spectroscopy, Raman scattering, and beyond
\cite{engel2007evidence,panitchayangkoon2010long}.

\subsection{Examples from chemistry}

The electronic structure formulation of quantum chemistry is a convenient approximation, but it is an incomplete description of the state of a molecule. Real molecules have not only electronic degrees of freedom, but also vibrational (and rotational) ones as well, i.e., fermi-boson mixed matter. As noted above, hybrid CV-DV quantum computing provides a natural paradigm to simulate these systems.

One unique characteristic in the chemistry setting is that that many vibrational modes in realistic molecules are anharmonic. This raises the question of how anharmonic quantum oscillators can be mapped to hybrid CV-DV quantum processors with only quantum harmonic oscillators. Interestingly, an entire framework of the algebraic theory of molecules \cite{iachello1995algebraic,van1983algebraic,iachello1982algebraic,iachello1991algebraic} has been developed in chemical physics that can effectively map a single anharmonic quantum oscillators into two or more harmonic oscillators, including both bound and scattering states. Examples include the common Morse \cite{alhassid1983group} vibrational modes for describing chemical bond breaking, the P\"oschl-Teller \cite{ arias20041} potential for chemical bond rotation, and other more general models \cite{barut1987algebraic}. The algebraic theory has been used in applications involving ro-vibrational spectra \cite{mccoy1992algebraic} and rotational-induced dissociation of molecules \cite{sondergaard2017rotational} in the classical computing setting. These theories also set solid foundations for quantum simulation of generic anharmonic CV degrees of freedom on hybrid CV-DV quantum processors.

In this section, we provide a few utility examples in chemistry where hybrid CV-DV quantum computing can be useful.

\begin{figure}[htbp]
    \centering
    \includegraphics[width=1.0\linewidth]{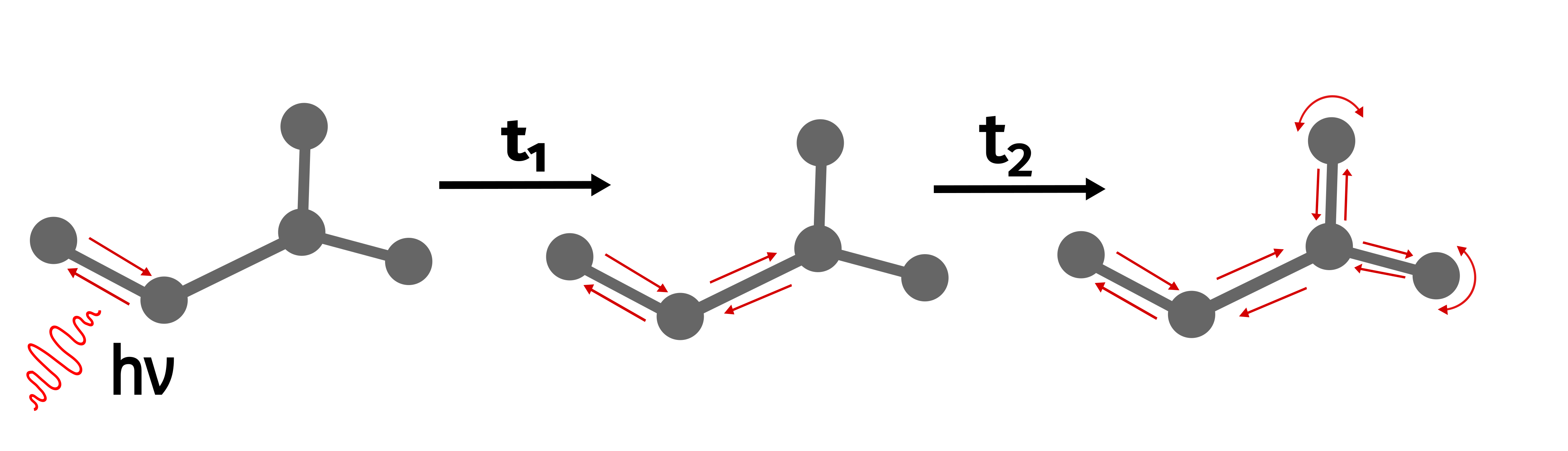}
    \caption{An illustration of the IVR process. A laser pulse excites a local vibrational mode, the energy of which is redistributed to the other vibrational modes over time.}
    \label{fig: IVR}
\end{figure}

\subsubsection{Intramolecular vibrational energy redistribution}

Intramolecular vibrational energy redistribution (IVR)\cite{boyall_modern_1997, nesbitt_vibrational_1996, karmakar_intramolecular_2020} refers to the study of the real-time evolution of the vibronic state of an isolated molecule after being put in some initially excited state $\ket{\psi(0)}$ using, for example, a laser pulse. The laser pulse initially excites a local mode, the energy from which will delocalize over time to a subset of the other vibrational modes of the molecule until the system thermalizes. An illustration of this process is given in Fig.~\ref{fig: IVR}, and involves bosonic couplings beyond the harmonic term:
\begin{align}
    H =  \sum_{j \neq k \neq l}^N \Phi_{jkl}(b_j^\dagger + b_j)(b_k^\dagger + b_k)(b_l^\dagger + b_l) + \cdots
    \label{eq: IVR Hamiltonian}
\end{align}

It is often the case in molecular chemistry that the rate and mechanism by which a reaction occurs (or whether it occurs at all) is state-dependent with respect to one or more of the involved chemical species. For example, a reaction may occur only if one of the reactants is in a particular geometric configuration, is ionized, or is in a particular electronic or vibronic excited state. By studying the time evolution of the bosonic coupling Hamiltonian, one can investigate whether or not the IVR process is responsible for or can be utilized to facilitate a particular reaction.

Similarly, one might know in advance that $\ket{\psi(0)}$ is an important intermediate transition state for some desired reaction. In this case, the quantity of interest is the survival probability $|\bra{\psi(0)}e^{-iHt}\ket{\psi(0)}|^2$. These survival lifetimes, which are molecule and mode-specific, are typically on the order of picoseconds to nanoseconds.~\cite{boyall_modern_1997} If the lifetime of $\ket{\psi(0)}$ is sufficiently long, then that reveals the possibility of performing bond-specific chemical reactions. That is, one designs a way of preparing a state $\ket{\psi(0)}$ with a specific locally excited vibronic mode, then carries out a chemical reaction for which that vibronic mode plays a key role. This is only possible if the lifetime of $\ket{\psi(0}$ is not short compared to the timescales involved in key steps of the reaction mechanism (\emph{e.g.} the average time between collisions of two molecular species or the time needed for a molecule to change geometric configurations).  The search for molecular systems with long IVR lifetimes is one of the main applications of IVR for this reason\cite{boyall_modern_1997}.

Classically, the time evolution %in Eq.~\ref{eq: IVR time evolution} 
can be emulated using quantum molecular dynamics methods such as the multiconfiguration time-dependent Hartree (MCTDH) method. Notably, the memory requirements of MCTDH scale exponentially with the number of degrees of freedom of the problem~\cite{meyer_quantum_2003}. Thus, much like exact diagonalization methods in electronic molecular chemistry, the use of such methods are limited to problem sizes with a modest number of degrees of freedom involved. MCTDH has, for example, been applied to study the role of IVR in the $trans$-$cis$ isomerization of formic acid in the gas phase~\cite{aerts_intramolecular_2022} and the dissociation of carbonyl sulfide~\cite{perez_quantum_2018}, which are comprised of 5 and 3 atoms, respectively.

One potential use-case of CV (or CV-DV) quantum computers therefore would be to supplant classical methods such as MCTDH that are highly accurate, but costly. Treating the vibronic degrees of freedom in the IVR problem requires having some information about the electronic potential surface of the molecule, either explicitly (conducting electronic calculations beforehand) or implicitly (time-evolving a Hamiltonian incorporating both electronic and vibronic degrees of freedom). Broadly speaking, there could be two classes of approaches to this use-case. The first would be to construct a workflow on  CV-DV processors wherein qubits efficiently compute estimations of the potential energy surface for a large number of degrees of freedom using an algorithm such as quantum phase estimation on-the-fly, while the oscillators carry out the time evolution of the nuclei Hamiltonian. This scheme does not require entanglement between the qubits and oscillators, and is therefore approximate and can be viewed as a version of \emph{ab initio} molecular dynamics accelerated on CV-DV quantum hardware.
Alternatively, since the IVR problem is fundamentally just the time evolution of a state comprised of fermionic and bosonic degrees of freedom coupled together, one could imagine performing the time evolution of a hybrid electronic-vibronic Hamiltonian, foregoing the necessity of calculating electronic potential surfaces altogether. Similar post Born-Oppenheimer approximation methods have been proposed for qubit quantum computers for calculating static properties of such systems.~\cite{veis_quantum_2016} Analogous methods could be proposed for calculating the dynamic properties relevant to IVR with the electronic and vibronic degrees of freedom treated on an equal footing. The benefit of using a CV-DV processor for such problems is that one does not incur the large overhead associated with the compilation of bosonic gates into qubit gates. 

\subsubsection{Vibronic Dynamics for energy harvesting}

The interplay of electronic and vibrational motion governs molecular light absorption, transfer, and emission, from fundamental dynamics to device performance. For instance, ultrafast S$_2$ $\to$ S$_1$ internal conversion in SO$_2$ can be rationalized through linear vibronic coupling, which captures the nonadiabatic character of the transition \cite{plasser2019highly}. The same principle underlies the tunability of chromophores such as BODIPY, where vibronic interactions determine how excitation energy is redistributed across molecular frameworks \cite{benniston2009lighting}. In optoelectronic systems, ligand design dictates exciton dynamics precisely because it alters vibronic pathways \cite{elik2023theoretical}, while in Ru(II) complexes, electroluminescence emerges only through the combined action of vibronic and spin–orbit couplings, collectively called the spin-vibronic effect~\cite{gao2000solid}. This effect has recently been experimentally studied in structurally related dinuclear Pt(II) metal–metal-to-ligand charge-transfer (MMLCT) complexes, which drives efficient singlet–triplet conversion \cite{Rather2023}. By probing vibronic coherences, it was observed that accelerated quantum-mechanically forbidden transitions at non-adiabatic crossings can happen. Vibronic coupling enables resonance between excitonic states and vibrational modes, sustaining coherence that enhances exciton migration \cite{doi:10.1021/acs.jpclett.5b02058}. Environmental dissipation, provided by protein scaffolds and solvent, prevents trapping of local excitations and enables “environment-assisted quantum transport” \cite{doi:10.1021/jz400058a}. In the Fenna-Matthews-Olson (FMO) complex, ultrafast spectroscopy confirms that vibronic coherence improves transfer efficiency under physiological conditions \cite{doi:10.1073/pnas.2112817118}. Together, these cases illustrate that vibronic effects are not secondary perturbations but the central mechanism by which molecular structure is translated into photophysical function.

In more detail, the dynamics can be described by the Lindblad equation, 
\begin{align}
    \frac{d\rho}{dt} = -\tfrac{i}{\hbar}[H_{\text{tot}}, \rho] + \mathcal{L}(\rho),
\end{align}
with Hamiltonian 
$H_{\text{tot}}=H_{\text{exc}}+H_{\text{vib}}+H_{\text{int}}+H_{\text{bath}}$, 
where $H_{\text{exc}}=\sum_i \epsilon_i a_i^\dagger a_i+\sum_{i\neq j}V_{ij}(a_i^\dagger a_j+\text{h.c.})$, 
$H_{\text{vib}}=\sum_i \omega_i b_i^\dagger b_i$, 
$H_{\text{int}}=\sum_i \lambda_i a_i^\dagger a_i(b_i^\dagger+b_i)$, 
and $H_{\text{bath}}=\sum_k \omega_k c_k^\dagger c_k+\sum_{i,k} g_{ik}a_i^\dagger a_i(c_k^\dagger+c_k)$. Key observables include site populations, coherences, exciton currents, sink efficiency, vibrational occupations, and experimentally accessible spectra and quantum yields \cite{engel2007evidence,panitchayangkoon2010long}. Quantum simulators have realized simplified Holstein and spin--boson models using trapped ions and superconducting circuits \cite{huh2017vibronic, RevModPhys.86.153}. Key algorithms include Trotter--Suzuki, qubitization, and quantum trajectory methods \cite{lloyd1996universal,low2017optimal}. Classical hierarchical equations of motion (HEOM) remains essential for benchmarking open-system dynamics \cite{10.1063/1.3155372}. Yet qubit-only methods face steep costs: bosonic modes require large truncations, Lindblad dynamics are not natively unitary, and non-Markovian baths with structured spectra are hard to encode \cite{RevModPhys.89.015001}. 

Hybrid CV--DV devices avoid this overhead by representing vibrations with oscillators and excitons with qubits, realizing native spin--boson couplings and engineered baths. Recently, schemes for simulating such nonadiabatic vibronic dynamics for photosynthetic chromophores on hybrid oscillator-qubit devices have been proposed, where environmental effects (dissipation, dephasing) were modeled with midcircuit measurements and resets \cite{vu2025computational,dutta2024simulatingchem}. A proposed hybrid CV-DV compiler utilizes generalized quantum signal processing (GQSP) algorithm to synthesize arbitrary bosonic phase gates, which then can decompose nonadiabatic molecular dynamics for uracil cations \cite{hong2025oscillatorqubitgeneralizedquantumsignal}.

\subsubsection{Chemical reaction dynamics through simulation of conical intersections}
Conical intersections (CIs) are degeneracies between electronic potential energy surfaces where the Born–Oppenheimer approximation breaks down due to strong electron–nuclear coupling. They govern ultrafast nonradiative transitions in photochemistry, proton-coupled electron transfer, and biomolecular processes such as vision and DNA photostability.  

Several diabatic Hamiltonians are widely used to model CIs. The most general is the multi-state linear vibronic coupling (LVC) or quadratic vibronic coupling (QVC) \cite{doi:10.1021/jp994174i, doi:https://doi.org/10.1002/9780470142813.ch2, MARTINEZ1997139} form, which may be written as in Eq.~\ref{eq:fermi-boson-hamiltonian} or in CV position/momentum coordinates (see App.~\ref{app:CI}).  In either case, the off-diagonal couplings generate the conical intersection.

CIs are characterized by a variety of measures. These include branching-plane vectors in the adiabatic representation: $\mathbf{g}=\nabla_{\mathbf{R}}(E_2-E_1)$ and $\mathbf{h}=(E_2-E_1)\,\mathbf{d}_{12}$ with $\mathbf{d}_{12}=\langle\phi_1|\nabla_{\mathbf{R}}\phi_2\rangle$, population transfer $P_{\alpha\!\to\!\beta}(t)$, branching ratios, quantum yields, nonadiabatic coupling norms $\|\mathbf{d}_{\alpha\beta}\|$, geometric phase signatures, and spectroscopic observables such as ultrafast pump–probe or 2D spectroscopy.

\subsection{Other examples from physics}

\subsubsection{Exciton-phonon coupling}

Phonons are key players in determining the physics of real materials.  While the primary focus is often on the electrons and their ground state,
phonons can and do have a significant effect on the properties of the material under study.  This is perhaps most clearly seen in conventional superconductors,
where (as shown by Cooper) there is a divergent instability in a pairing channel mediated by phonons; this observation led to the BCS theory of
superconductivity, and many experimental confirmations that phonons are indeed the pairing glue between electrons.  But, even outside this
particular case, phonons show up across the spectrum.  This can be in the development of ordered phases such as charge density waves~\cite{gruner1994dynamics}
but also in more mundane places such as metals, where phonons are responsible for a characteristic temperature dependence of the resistivity~\cite{abrikosov2017fundamentals}.

While the effects of phonons are well understood in certain contexts, they retain some mystery in others. One such context is the
transport properties of organic and hybrid semiconductors, where there is strong coupling to the soft modes of the relatively low weight atomic groups. 
One
notable example
is transport through organic and hybrid thin film transistors~\cite{orgiu2015conductivity, de2017vibronic,deng2025organic,horowitz2004organic}.  This is a problem with potentially high societal impact, yet many questions
remain regarding the origin of the limitations of the photocurrent; in turn, this prevents informed device design that overcomes these limitations.
It is thought that phonons play a key role, but in addition the photon modes that couple to the excitations in the material need to be considered. 
With current techniques, we have no fully-quantum model that can handle the full process --- from initial excitation to steady state transport --- and
are thus limited in developing a holistic understanding of this problem.

Phonons may be coupled to the excitons (and indeed to electrons as well) in a variety of ways~\cite{mahan,bruus}. Two typical forms are the Holstein
and Peierls (or Su-Schrieffer-Heeger\cite{su1979solitons}) couplings, which may be written as~\cite{fetherolf2020unification}
\begin{align}
    \mathcal{H}_{\text{Holstein}} &= g \sum_i \delta X_i c^\dagger_i c_i \\
    \mathcal{H}_{\text{Peierls}} &= g \sum_{\langle i,j \rangle} (\delta X_j - \delta X_i) c^\dagger_j c_i.
\end{align}
In the Holstein case, the displacement of the ionic position $X_i$ causes a local change in the exciton energy; in this case,
a transfer integral is modulated between nearest neighbor exciton sites.  As is typical, the local displacement may be written as $X_i = b^\dagger_i + b_i$,
which reveals the bosonic operators that need to be taken care of, and the potential for CV/DV methods to play a role.

With the underlying Hamiltonian established, we can now discuss the most relevant observables.  The most fundamental quantity governing the device properties is the photocurrent, which arises from a combination of the excitation rate, the effective dissipation rate, and the transport characteristics of the material. This quantity can be initially examined through the optical conductivity, obtained from the current–current correlation function:
\begin{align}
\text{Re}\ \sigma(q,\omega) = \frac{1}{\omega V} \int_0^\infty dt e^{i\omega t}\ \mathrm{tr}\left[ \rho j^\dagger(q,t) j(q,0) \right].
\end{align}
Here, $j$ denotes the current operator
$j(q) = \sum_k c^\dagger_{k+q} c_k$.

\subsubsection{Bose-Hubbard models}

\begin{figure}
    \centering
    \includegraphics[width=1.\linewidth]{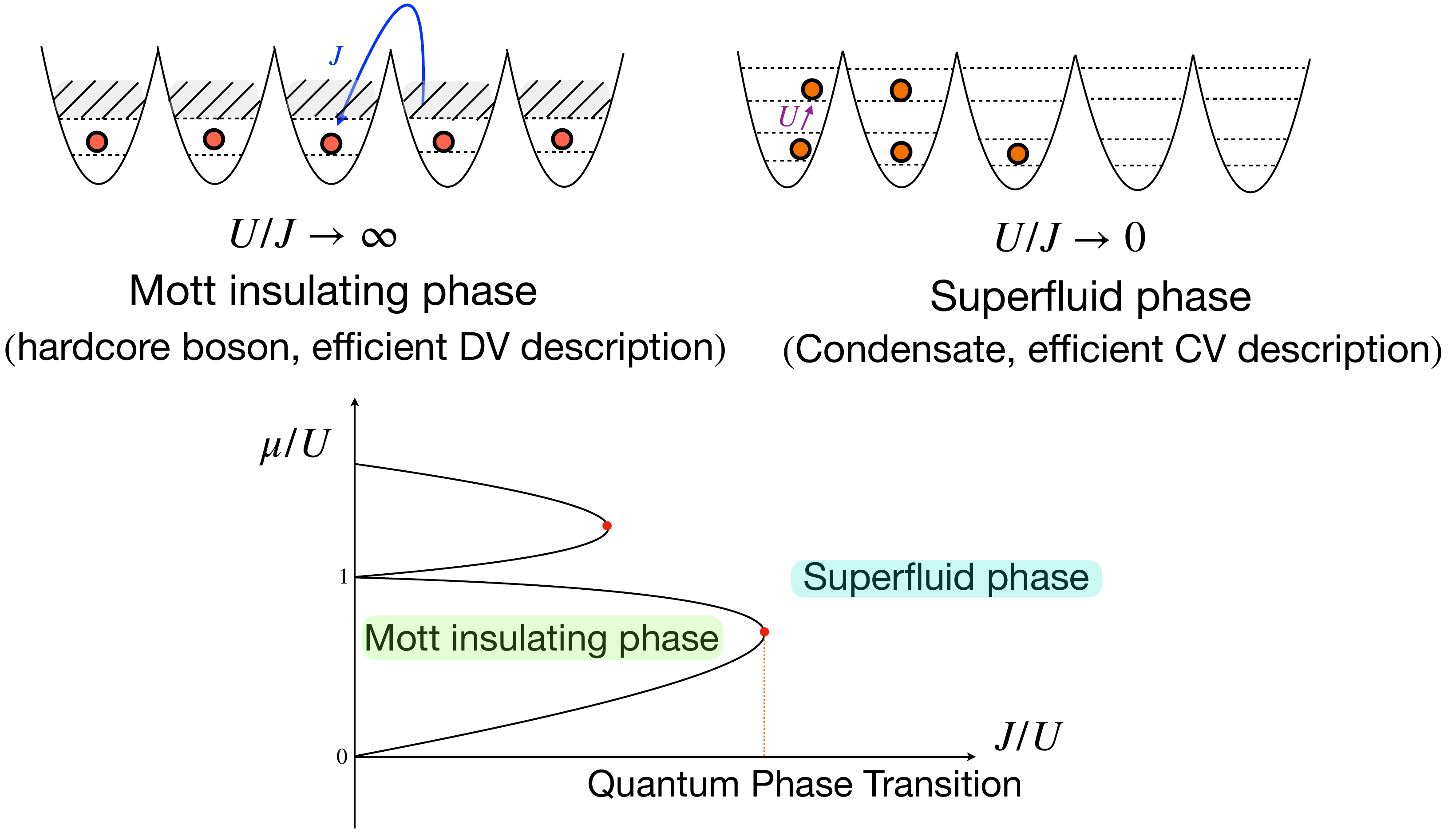}
    \caption{The two extreme limits of the Bose-Hubbard model with $N = L = 5$ (unit filling). In the superfluid phase, depending on the filling and the ratio $U/J$ all the bosons can occupy (condense) to a particular lattice site. In the lower panel, we show the conjectured phase diagram at zero temperature where the tip of Mott lobes (red circles) correspond to the BKT-type phase transition found in the 2d classical XY model. The particle density is fixed for a given lobe and changes by $1$ as we increase $\mu$.}
    \label{fig:bose-hubbard1}
\end{figure}

One particularly useful example where the efficient description in terms of discrete and continuous variables interpolates between two limits is the Bose-Hubbard model. 
A few previous studies of this model in 1+1 dimensions have explored its real-time dynamics~\cite{Kalajdzievski2018}, while others have employed variational methods to approximate the ground-state energy over a range of 
$U$ values~\cite{Yalouz:2021oyo, Stornati:2023kku}. 
In a specific limit where efficient qubit encoding is possible, this model was also studied in 2+1-dimensions on a superconducting qubit device~\cite{Yanay2020}. The Bose-Hubbard Hamiltonian is given by:
\begin{align}
\label{eq:BH_Hamiltonian}
    H  = &-J \sum_{j=1}^L
    \left( \hat{a}_j\hat{a}_{j+1}^{\dagger}+ \text{h.c.} \right)
    \pm \frac{U}{2} \sum_{j=1}^L \hat{n}_j\left( \hat{n}_j-1 \right) \nonumber\\ &- \mu \sum_{j} n_{j},
\end{align}
where $\hat{a}, \hat{a}^\dagger$ are the bosonic operators, $-(+)$ represents the attractive (repulsive) versions of the model, $U$ is a positive real parameter, and h.c. denotes the Hermitian conjugate. The attractive version is numerically more challenging than the repulsive Bose-Hubbard model. The parameter $J$ is known as the `hopping parameter' while $U$ is known as the on-site term. The Hamiltonian is invariant under global $U(1)$ symmetry i.e., $\hat{a}_{j} \to \hat{a}_{j} e^{i\phi}$. When the on-site repulsion is zero i.e., $U=0$, then we have just the hopping term and it becomes a tight-binding model. If $U/J \to \infty$, then the bosons behave in a manner different from a free boson and we call them `hardcore bosons', avoiding the occupation of the same cavity (or lattice site). Therefore, while the large $U$ limit can be efficiently studied using DV, for small values of $U$ and particularly in the superfluid phase, an approach based on CV is likely more efficient. This is the Mott insulator phase (named due to the lack of conductance). When we go to the opposite limit, i.e., $U/J \to 0$, then we can have all bosons in a particular cavity, which is referred to as the superfluid phase. This model exhibits the Mott insulator (MI)–to–superfluid (SF) transition first studied in Ref.~\cite{Fisher1989}, which was later observed experimentally in cold-atom systems confined to optical lattices~\cite{Greiner2002}, where the quantum phase transition between the two phases was realized at temperatures close to absolute zero. It would be interesting to study this model using CV methods and follow the quantum phase transition and also study out-of-equilibrium properties~\cite{Kennett2013}.
In Fig.~\ref{fig:bose-hubbard1}, we show the phase diagram of this model and also how the MI phase has a restricted occupancy while in the superfluid phase, all the bosons $N$
can effectively condense at a particular site, rendering the qubit encoding less effective. 

\subsection{Quantum field theories}

\begin{figure*}[htpb]
    \centering
    \includegraphics[clip=true,trim=30 0 20 0,
    width=0.99\textwidth]{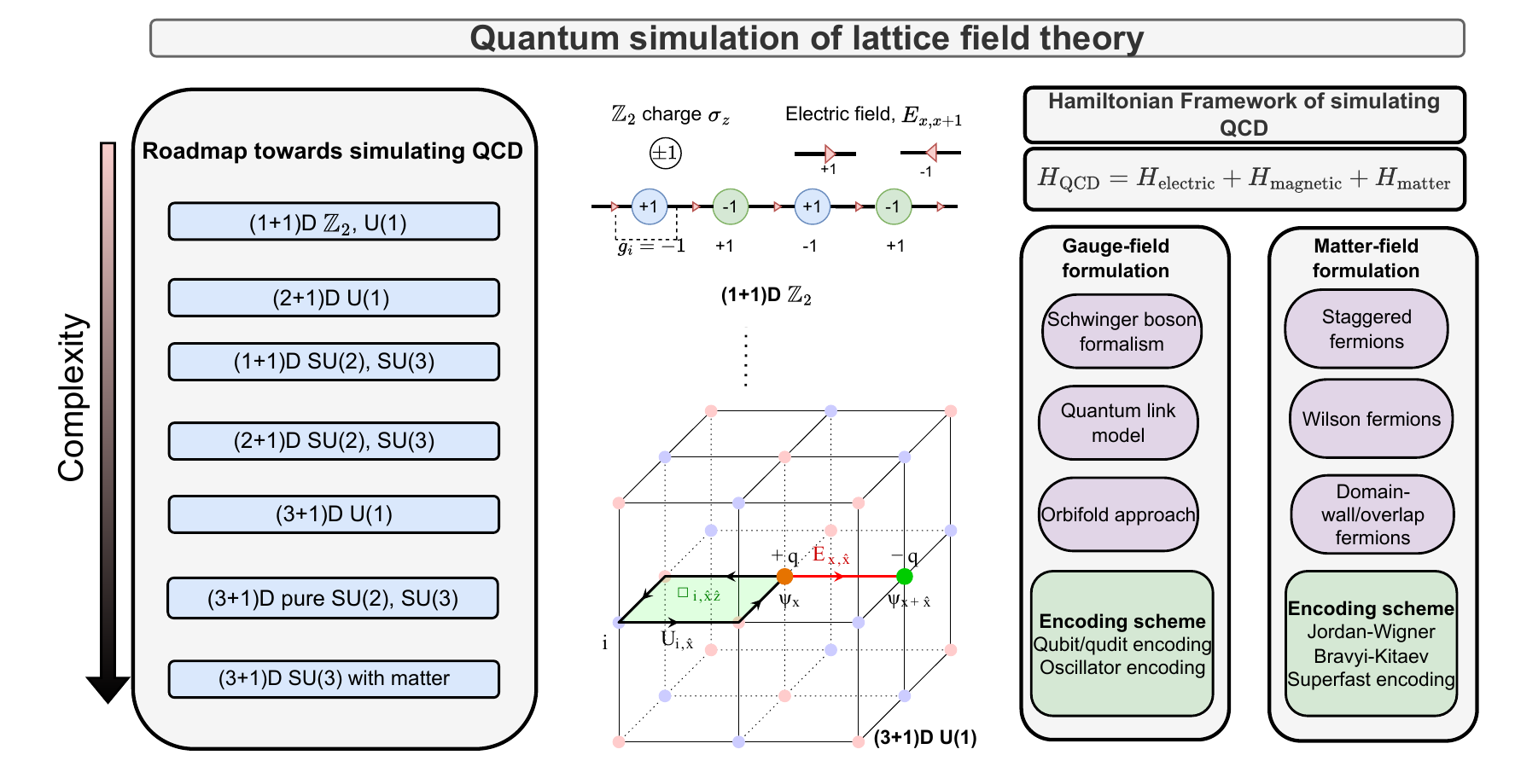}
    \caption{Inspired by the sequence of theories in \cite{kogut1979introduction}, we show a roadmap towards simulating QCD with quantum simulators (left). The complexity scale presented is not rigorous, rather it identifies relative algorithmic difficulties and quantum resource requirements in implementing respective theories. Representative field content for generic field-theoretic constructions in (1+1)D and (3+1)D are shown (middle). An example field configuration is shown in a particular gauge-invariant sector identified by $\{g_i\}$ (middle-top) for the $\mathbb{Z}_2$ gauge theory. Magnetic term (green), electric term (red), matter field ($\psi_x$) on vertex and gauge fields $U_{x,\mu}$ on edges are identified (middle-bottom). The quantum simulation framework for lattice field theories and some of the approaches to encode gauge fields and matter fields are identified (right).}
    \label{fig:lattice_roadmap}
\end{figure*}

One of the challenges in simulating quantum field theories involving bosonic or gauge degrees of freedom is the infinite-dimensional local Hilbert space. This requires truncation before they can be studied using either classical or quantum computers. However, quantifying the impact of the truncation is often a non-trivial problem.
There are different lattice descriptions that can lead to the desired continuum description, however, maintaining a proper balance between truncation and extracting relevant physics in the continuum limit is a hard problem. In addition, one might have a good truncated lattice model that reproduces correct continuum physics, but if the resources are exponential in system size then it is not useful. Therefore, it is essential to seek a lattice gauge theory formulation with a finite local Hilbert space that reproduces the correct continuum physics while requiring only moderate computational resources.

In quantum field theories, we have gauge fields, scalar fields, and fermion (matter) fields. While it is efficient to use qubits to encode the fermions, for gauge and scalar fields, this might not be the most efficient method. 
Rather, an alternative approach is to encode the scalar and gauge fields in terms of continuous variables. This can be a powerful method for simulating lattice gauge theories where fermionic fields can be represented by discrete variables and scalar and gauge fields can be represented by continuous variables~\cite{Lloyd:2000bwc, Andersen:2014xqr, Davoudi:2021ney, Sutherland2021, Liu:2024mbr, crane2024hybrid, Araz:2024dcy, Ale:2024uxf, Than:2025gso,girvin2025toward}. 
The long-term goal is to study quantum field theories with local gauge invariance in 3+1-dimensions describing fundamental interactions in nature. We show a roadmap towards this goal in Fig.~\ref{fig:lattice_roadmap}.  

The first step in studying models with continuous symmetries—whether global or locally gauged—is to address the problem of truncation, i.e., constructing an efficient finite local Hilbert space formulation.
Over the past decades, several strategies have been explored to achieve this goal. These include the finite subgroup or group-space decimation approach~\cite{rebbi1980phase, flyvbjerg1984group, hasenfratz2001asymptotic, Alexandru:2021jpm, Alexandru2022, Gustafson:2024kym}, fuzzy regularization~\cite{Alexandru:2023qzd}, the quantum groups method~\cite{Zache:2023dko, Hayata:2023bgh}, and direct truncation of the continuous group~\cite{Bazavov:2019qih, Meurice:2020pxc, Illa:2024kmf, Balaji:2025afl}.

Other notable approaches include the quantum link model~\cite{Horn1981, Orland1990, Chandrasekharan:1996ih, Brower:1997ha, Wiese:2013uua}, where gauge invariance is maintained but the quantum links become nonunitary—unitarity being recovered only in the $S \to \infty$ limit—and the Schwinger boson or prepotential-inspired loop–string–hadron (LSH) approach~\cite{Mathur:2000sv, Mathur:2004kr, Anishetty:2009nh, Anishetty:2009ai, Raychowdhury:2019iki, Kadam:2022ipf, Yang:2025edn}, which constructs gauge-invariant states by expressing gauge links in terms of $SU(2)$ or $SU(3)$ Schwinger bosons.
Additional formulations include the orbifold lattice approach~\cite{Bergner:2024qjl} and the $1/N$ expansion in the planar-limit framework~\cite{Ciavarella:2024fzw}.
However, it is not clear which of these approaches would eventually lead to the most promising route for future simulations of quantum field theories relevant for strong interactions on fault-tolerant quantum computers. In addition, there has also been several works discussing effects of truncation in quantum simulation of lattice gauge theories for a given accuracy~\cite{Tong:2021rfv, Kan:2021xfc}. However, the relation between the truncation effects and error scaling as we take the continuum limit remains an open problem for $SU(2)$ and $SU(3)$ lattice gauge theories.
We emphasize that encoding bosonic and fermionic quantum fields using a CV--DV approach might be essential in scaling and ensuring that correct continuum physics is obtained. Several previous works have made use of this for the simulation of quantum many-body systems and quantum field theories~\cite{Marshall:2015mna, Yang:2016hjn,Thompson:2023kxz,Davoudi:2021ney, Liu:2024mbr,crane2024hybrid,Araz:2024dcy}.

\subsubsection{O(3) sigma model as rigid rotor}

The simplest non-trivial example of a quantum field theory is the scalar field theory. This has been 
extensively studied in 1+1-dimensions from a quantum computation perspective~\cite{Jordan:2011ci,Jordan:2017lea, Hardy:2024ric} including continuous variable method~\cite{Thompson:2023kxz,Abel:2024kuv}. As a pathway to the future fault-tolerant computation of non-Abelian lattice gauge theories, including 3+1-dimensional quantum chromodynamics (QCD), a particularly interesting example is the $O(3)$ non-linear sigma model~\cite{Polyakov1975, Migdal:1975zf} in 1+1-dimensions. This is also natural since the Kogut-Susskind Hamiltonian we will discuss later has deep connections to the rotor formulation of the $O(3)$ model. Though this is not a gauge theory, it has a non-Abelian global compact $O(3)$ symmetry and can be expressed completely in terms of a Hamiltonian of a rigid rotor as first proposed by Hamer, Kogut, and Susskind~\cite{Hamer:1978ew} within the broad class of $O(N)$ models. 
This model shares several features with QCD, such as being asymptotically free for $N \ge 3$~\cite{Polyakov1975}. It serves as an interesting toy model because it avoids the additional complexities associated with local gauge degrees of freedom while retaining nontrivial dynamics due to its non-Abelian global symmetry. 
The lattice Hamiltonian on a one-dimensional spatial lattice of $N$ sites is given by: 
\begin{equation}
    H=\frac{1}{2 \beta} \sum_{i=1}^N \mathbf{L}_i \cdot \mathbf{L}_i -\beta \sum_{\langle i j \rangle} \mathbf{n}_i \cdot \mathbf{n}_j,
\end{equation}
where $\mathbf{L}_i$ is the angular momentum operator and $\mathbf{n}_i$ is a unit 3-vector that takes values on the two-sphere $S^2$. The coupling $\beta$ determines the interaction and the continuum $O(3)$ field theory is obtained as we take $\beta \to \infty$. 
This model has been studied using qubit regularization and the fuzzy approach, where it was found that, instead of requiring an infinite-dimensional local Hilbert space, a four-dimensional local space is sufficient to reproduce the correct continuum theory at the asymptotic fixed point as $\beta \to \infty$~\cite{Singh:2019uwd, Bhattacharya:2020gpm, Alexandru:2021jpm}.
In addition to the discrete-variable approach, this model has also been studied using continuous variables by recasting the rotor Hamiltonian in terms of Schwinger bosons or triplet scalar fields with a unit-norm constraint corresponding to the $O(3)$ symmetry~\cite{Jha:2023ecu, Jha:2023ump}.
It would be interesting to understand the higher-dimensional $O(N)$ rotor models using a CV approach.

%%%%%%%%%%%%%%%%%%%%%%%%%%%%%%%%%%%%
% References from her onward checked
%%%%%%%%%%%%%%%%%%%%%%%%%%%%%%%%%%%%
\subsubsection{Abelian and non-Abelian lattice gauge theories}

The simplest lattice gauge theory (LGT) is the $\mathbb{Z}_{2}$ model~\cite{Wegner:1971app, Mildenberger:2022jqr}. However, we begin with the $U(1)$ Abelian gauge theory, as it represents the simplest case that is also part of the Standard Model of particle physics. 
The $U(1)$ gauge theory has been extensively studied in 1+1 dimensions, where the entire Hamiltonian can be expressed in terms of discrete variables by integrating out the gauge degrees of freedom~\cite{Muschik:2016tws, Farrell:2024fit}. It has also been explored in 2+1 dimensions, where an efficient discrete representation in terms of quantum links is possible~\cite{Gonzalez-Cuadra:2024xul}.
 
For the case of $U(1)$ gauge theory, let us start with the Kogut-Susskind (KS) Hamiltonian~\cite{Kogut:1974ag, Zohar:2014qma}, developed soon after the Lagrangian formulation of non-Abelian lattice gauge theories by Wilson~\cite{Wilson:1974sk}. In a Hamiltonian formulation~\cite{Dirac1949} of LGT, we keep the time continuous and discretize the $d$ spatial directions leading to $(d+1)$-D quantum field theory. In general, the \emph{pure} gauge KS Hamiltonian consists of two terms: electric, and magnetic given by Eqs.~(\ref{eqn_electric}–\ref{eqn_Hmagnetic}). The magnetic term is only present for $d > 1$ since there is no $\vec{B}$ in one (spatial) dimension. To generate non-trivial gauge dynamics, the two terms in the Hamiltonian must not commute with each other and the terms in the KS Hamiltonian satisfy this property. Therefore, it is not possible to find a basis where both terms are diagonal. The choice to make either the electric term or the magnetic term diagonal is referred to as the `electric’ and the `magnetic’ basis. The electric basis is not very efficient for obtaining the continuum limit. Rather than the electric basis, we can use the magnetic basis~\cite{Romiti:2023hbd} where the unitary links are diagonal or a mixed electric/magnetic basis~\cite{DAndrea:2023qnr} as has been pursued in recent papers.  
The Hamiltonian of a general LGT with matter for the case of $U(1)$ gauge theory can be written as 
\begin{equation}
H = H_{\mathrm{electric}} + H_{\mathrm{magnetic}} + H_{\mathrm{matter}},
\label{eqn_matter_gauge}
\end{equation}
where each term naturally maps onto either qubit or qumode (continuous) degree of freedom in a hybrid encoding. The electric field $E_{\mu}(\textbf{n})$ (starting at site $\textbf{n}$ and oriented in the $\hat{\mu}$ direction) has an infinite integer spectrum and is defined on all links of the lattice and is naturally represented by a qumode quadrature. The electric and magnetic terms are:
\begin{equation}
H_{\mathrm{electric}} = \frac{g^2}{2} \sum_{\textbf{n},\,\mu=1, \cdots d} E_{\mu}^2(\textbf{n})
\label{eqn_electric},
\end{equation}
\begin{equation}
H_{\mathrm{magnetic}} = -\frac{1}{2g^2} \sum_{\square} \rm{Tr} \left( U_{\square} + U_{\square}^\dagger \right), \label{eqn_Hmagnetic}
\end{equation}
where $U_{\square} = U_{\mu}(\textbf{n}) U_{\nu}(\textbf{n}+\hat{\mu}) U^\dagger_{\mu}(\textbf{n}+\hat{\nu}) U^\dagger_{\nu}(\textbf{n})$ defined as the product of unitary link operators around a given plaquette (Wilson operator) with origin at $\textbf{n}$. This can be naturally encoded in terms of qumodes with the link at site $\textbf{n}$ in direction $\mu$ defined as
$U_{\mu}(\textbf{n}) = e^{ig A_{\mu}(\textbf{n})}$ with lattice spacing $a_s=1$. The electric field and gauge link satisfy $[E(x), U(x)] = U(x)$ or alternatively, $[A(x), E(x^\prime)] = i\delta_{x x^{\prime}}$ where $U(x)$ denotes the gauge link starting at site $x$ in either of the $d$ directions. 

The fermionic matter term in the lattice Hamiltonian requires a specific discretization scheme, which is nontrivial due to the fermion doubling problem that arises under a naive lattice discretization.
To overcome this, there are various proposals and one can choose either staggered~\cite{Susskind:1976jm}, Wilson fermions~\cite{Wilson:1974sk, Ginsparg1982}, domain-wall~\cite{Kaplan1992}, or overlap fermions~\cite{Neuberger:1997fp}. In addition, we will only consider one fermion flavor ($N_f = 1$). The matter term is given by:
\begin{align}
    H_{\mathrm{matter}} &=  \kappa \sum_{\textbf{n},\mu} \Big( \eta_{\mu}(\textbf{n}) \,\psi(\textbf{n})^\dagger U_{\mu}(\textbf{n}) \psi({\textbf{n} +a_{s}}) 
+ \mathrm{h.c.} \Big) + \\ 
& m\sum_\textbf{n} (-1)^{\sum_\mu n_\mu} \psi(\textbf{n})^\dagger \psi(\textbf{n}) \nonumber,
\end{align}
where $\kappa$ is the matter-gauge coupling, $\text{h.c}$ denotes the Hermitian conjugate, $\eta_{\mu}$ is a site-dependent scalar phase factor encoding the structure of $\gamma$-matrices, and $\mu = 1, \cdots, d$. In the staggered formulation, the upper (lower) components of the spinor are denoted by fermionic fields on different neighboring (odd and even) sites. The fermions have a finite-dimensional local Hilbert space and are encoded in qubits using Jordan-Wigner (JW), Bravyi-Kitaev (BK)~\cite{Bravyi:2000vfj}, or generalized superfast encodings~\cite{Setia:2018svm}. The fermionic operators act on qubit registers, while the electric field $E_{\mu}({\textbf{n})}$ and link operators $U_{\mu}(\textbf{n})$ can be encoded in CV degrees of freedom.
Finally, we have the Gauss's law condition (for example for $U(1)$ gauge theory) which requires both CV and DV degrees of freedom defined as 
\begin{align}
G(\textbf{n}) = \sum_{\mu = 1}^{d} \Big[ E_{\mu}(\textbf{n}) - E_{\mu} (\textbf{n} - a_s))\Big] \nonumber \\
- 
\big(\underbrace{\psi^\dagger(\textbf{n}) \psi(\textbf{n}) - \frac{1 - (-1)^{\sum n_i}}{2}}_{Q(\textbf{n})}\big),
\label{eq:gauss-law}
\end{align}
at each lattice site $\textbf{n} = (n_1, n_2, n_3)$. The local charge operator for the staggered fermion case, $Q({\textbf{n}})$, acts on qubits and the electric fields $E_{\mu}({\textbf{n})}$ act on oscillator/bosonic degrees of freedom. This constraint acts at each site $\textbf{n}$ and physical states $\ket{\psi}$ are those that satisfy $G({\textbf{n}}) \ket{\psi} = 0 \, \forall \, \textbf{n}$.

\begin{center}
\vspace{1em}
\begin{table}[h!]
\begin{tabular}{c|c}
\textbf{Term} & \textbf{Efficient encoding}\\
\hline
Electric field term $E_{\mu}^2({\textbf{n})}$ & CV \\
Magnetic (plaquette) term & CV \\
Pure fermionic terms ($\in H_{\mathrm{matter}}) $& DV \\
Gauge-matter term & DV+CV \\
Gauss's law constraint (Eq.~\eqref{eq:gauss-law}) & DV+CV \\
\end{tabular}
\caption{A schematic description of how different terms in the $U(1)$ Kogut-Susskind Hamiltonian LGT can be encoded in the CV and DV variables.}
\end{table} 
\end{center}

\begin{figure}[htpb]
    \centering
    \includegraphics[width=0.5\textwidth]{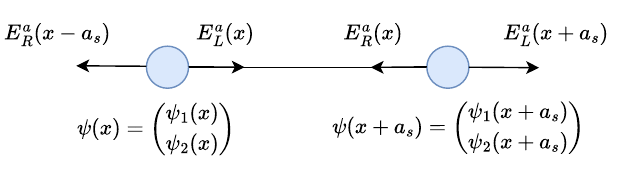}
    \caption{The electric fields at two lattice sites (with lattice spacing $a_s$) are shown with the two-component fermionic fields at each site for the $SU(2)$ Hamiltonian LGT in 1+1-dimensions.}
    \label{fig:su2-1d}
\end{figure}
Let us consider the case of $SU(2)$ as an example in one spatial dimension where $\textbf{n}$ is just $x$. For this case, Kogut and Susskind argued to consider a single non-Abelian gauge link as a rigid rotor such that left and right electric fields, denoted $E^a_{L}(x)$ and $E^a_{R}(x)$ respectively (see Fig.~\ref{fig:su2-1d}), correspond to the left and right end of a link for a given color index $a$, and we have $E_{L}^{2} = E_{R}^{2} = (E^{1}_{L})^2 + (E^{2}_{L})^2 + (E^{3}_{L})^2$. The left and right electric field satisfy, 
$[E_{L}^{a}, U] = \frac{1}{2}T^a U$ and
$[E_{R}^{a}, U] = U \frac{1}{2}T^a$
where $T^a$ are the three generators (Pauli matrices) of $SU(2)$. The magnetic term will be the trace of the oriented product of four $SU(2)$ links. The matter term will involve $\psi(\textbf{n})$ at each site with two $SU(2)$ color components. 

In Ref.~\cite{Ale:2024uxf}, the \emph{pure} $SU(2)$ gauge theory was studied in 2+1 dimensions within the maximal tree gauge~\cite{Creutz:1976ch}, using continuous variables by recasting the problem in terms of rigid rotor dynamics on the three-sphere $S^{3}$~\cite{Chin:1985ua}.
However, the primary goal of quantum simulations of lattice field theory is to study the theory of strong interactions (QCD), which is based on an $SU(3)$ gauge symmetry with a sufficient number of fermion flavors, $N_f$.
In coming years, it would be interesting to find a description for $SU(2)$ and $SU(3)$ gauge theory with matter using hybrid CV--DV approaches. 
In addition, 
it would be interesting to use trapped-ion or circuit quantum electrodynamics (cQED) platforms to simulate nontrivial dynamical phenomena, such as hadron scattering and string breaking, in these quantum field theories.

While the main focus is on quantum field theories relevant to the fundamental interactions of nature, such as quantum electrodynamics (QED) and quantum chromodynamics (QCD), the hybrid CV--DV framework also offers a promising avenue for exploring field theories beyond the Standard Model (BSM), which involve a rich interplay between bosonic and fermionic degrees of freedom.
One well-known class of such QFTs are the supersymmetric field theories such as supersymmetric quantum mechanics~\cite{Witten:1982im} and Wess-Zumino-type~\cite{Wess1974} models. In addition to these models, their extensions to maximally supersymmetric gauge theories~\cite{Catterall:2017lub, Catterall:2020nmn} are expected to play a crucial role in quantum gravity, especially through the AdS/CFT correspondence~\cite{Maldacena:1997re}. 

Hybrid CV--DV platforms can be an exciting avenue for such explorations extending some preliminary investigations~\cite{Crichigno:2020vue, Culver:2021rxo} using qubit-only approaches.
With current classical computing approaches, it is not possible to study real-time dynamics of these models and would be a excellent application of future fault-tolerant quantum computing and would provide insights into various non-perturbative aspects of quantum gravity.

%%%%%%%%%%%%%%%%%%%%%%%%%%%%%%%%%%%%%%%%%%%%%%%%%%%%%%%

\section{Computing-Relevant Applications}
\label{sec:other-computing}

In this section, we survey and discuss other applications that CV-DV quantum hardware can potentially address beyond quantum simulation of physical sciences. These problems include non-local games (Sec.~\ref{ssec:games}), optimization (Sec.~\ref{ssec:opt}), factoring (Sec.~\ref{ssec:factoring}), data compression and communication (Sec.~\ref{ssec:compress-comm}), differential equation solver (Sec.~\ref{ssec:diffq}), and quantum sensing (Sec.~\ref{ssec:sensing}). We comment on the potential of hybrid CV–DV hardware to achieve fault tolerance in Sec.~\ref{ssec:ft}, including new avenues for robust CV quantum computing.  

\subsection{Non-local games}
\label{ssec:games}

In quantum information, a game refers to a structured test, called a nonlocal game, that evaluates how well a quantum system can demonstrate correlations or behaviors that are impossible for classical systems. Games can provide an ideal benchmark platform to test CV-DV hybrid systems capabilities due to their minimal circuit depth requirements and ability to demonstrate provable
quantum advantage~\cite{Daniel_2022,Hart_2025,furches2025application}.  Unlike traditional quantum algorithms requiring fault-tolerant
architectures, nonlocal game protocols can operate effectively on near-term devices while providing rigorous certification of quantum resources. Recent
work has demonstrated their utility as application-level benchmarks using variational methods to successfully generate
short-depth quantum strategies for complex games including graph coloring problems~\cite{furches2025application}.

For CV-DV platforms, several promising directions emerge. First, games with large output spaces naturally favor CV systems, as
continuous measurements via homodyne detection provide access to unbounded outcome sets unavailable to discrete systems. The CHSH inequality has been
experimentally violated using CV states with four modes and homodyne detection~\cite{Thearle_2018},
demonstrating that squeezed states and continuous measurements can generate nonlocal correlations comparable to entangled qubits. The variational
approach of Furches et al.~\cite{furches2025application} for finding optimal quantum strategies could potentially be extended to hybrid CV-DV systems,
where the Hamiltonian encoding game rules would incorporate both discrete and continuous degrees of freedom.

Cat states offer unique advantages for hybrid nonlocal strategies. The superposition structure $|\alpha\rangle + |-\alpha\rangle$ naturally maps to
binary outcomes while maintaining CV resource advantages, and their non-Gaussian character has been shown to enable Bell
violations~\cite{vlastakis2015characterizing}. 
Recent work has demonstrated that variational quantum algorithms can discover optimal quantum strategies
for the Magic Square Game, achieving perfect game performance~\cite{chehade2025gametheoreticquantumalgorithmsolving}. Building on these variational approaches, we believe that cat state encodings could provide alternative implementations for games like the Magic Square Game, potentially offering advantages in
noise resilience due to the error correction properties of cat codes, while maintaining the perfect winning probability achievable with discrete
strategies.

The resource efficiency of nonlocal games makes them particularly suitable for benchmarking CV-DV interfaces. A typical protocol requires: (i)
preparation of entangled CV-DV pairs, (ii) local measurements, and (iii) classical post-processing to verify winning conditions. This simplicity contrasts sharply with other
algorithms requiring hundreds of circuit evaluations or error correction protocols demanding extensive syndrome extraction. The benchmarking methodology
of Furches et al.~\cite{furches2025application}, which tested strategies across 14 different quantum platforms, provides a template for systematic
CV-DV hardware evaluation using win rates as a noise-sensitive metric directly correlated with gate fidelity and bit-flip errors.

Furthermore, some nonlocal game strategies provide device-independent certification~\cite{supic2020self}. Violation of Bell inequalities guarantees the presence of
entanglement without assumptions about the internal workings of the devices, making them robust benchmarks for heterogeneous CV-DV architectures where
precise calibration across different physical platforms remains challenging. Furthermore, work on loophole-free Bell tests using high-efficiency homodyne
detection~\cite{garcia2005loophole} has paved the way for robust CV-DV implementations, while the graph coloring game with quantum
advantage~\cite{furches2025application} suggests opportunities for encoding graph properties in CV-DV phase space.

\subsection{Optimization}
\label{ssec:opt}

\subsubsection{Quadratic Unconstrained Binary Optimization under Fock Basis}

Variational quantum algorithms (VQAs) are a promising approach to
solve relevant quantum chemistry (Variational Quantum Eigensolver -
VQE) and optimization problems (Quantum Approximate Optimization
Algorithm - QAOA) algorithmically on quantum computers~\cite{Peruzzo2014,farhi2014quantum}.
Repeated iterations of VQAs alternate between execution of a
parametrized quantum kernel and classical optimization on the output
of the kernel to update parameters. Most state of the art quantum
devices (superconducting, ion traps, annealers, etc.) that can execute
circuits of these kernels are limited to qubit-based computations, but
the use of qumode-based computations is rapidly growing.  While there
are some known instances of problem agnostic kernels for hybrid
devices~\cite{eickbusch_fast_2022}, implementing problem-aware ansatzes on
either exclusively qumode based devices or hybrid devices warrant
further investigation.

\begin{figure}[htb]
    \subfigure[
        \label{subfig:maxcut} Maximum Cut]
        {\includegraphics[width=0.4\linewidth,clip]{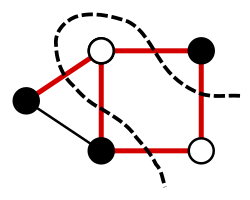}}
    \subfigure[
        \label{subfig:minvert} Minimum Vertex Cover]
        {\includegraphics[width=0.5\linewidth,clip]{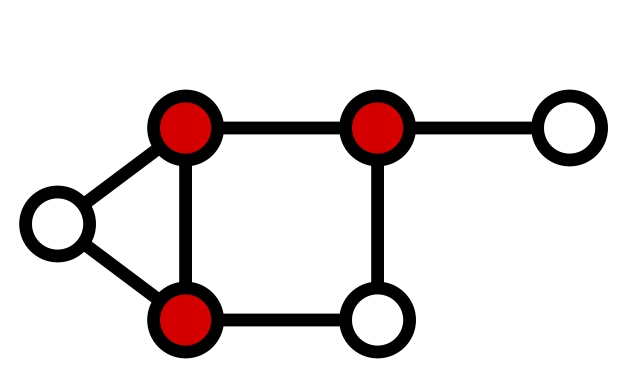}}
    \caption{Graph problems which can be expressed as QUBOs. (a) shows the cut (dotted line) made to partition the vertices into two disjoint sets (black and white).  It is considered a maximum cut because it maximizes the number of edges between the two sets. (b) shows the 3 vertices (red) which make up the minimum vertex cover. It is considered a vertex cover because all edges are incident on at least one of the red vertices, and is minimum because a vertex cover cannot be made with fewer red vertices.}
    \label{fig:optimization_example_problems}
\end{figure}

Therefore, we aim to compare the efficacy of qumode and hybrid devices
as compared to exclusively qubit based devices.  For straightforward
comparisons, we limit ourselves to a class of discrete-valued
problems, those which can be represented as Quadratic Unconstrained
Binary Optimization (QUBOs), 
\begin{equation}
  \underset{x\in \mathbb{B}^n}{\text{Minimize}} \; x^T Q x ,
\end{equation}
some examples of which are included in \cref{fig:optimization_example_problems}.
QUBOs can also be directly transformed into an Ising spin Hamiltonian by a change of variables,
namely $ \displaystyle{x = \frac{s+1}{2}}$.  Further details can be found in \cite{dwaveocean}.  

These problems naturally map to qubit-based devices and can somewhat
trivially map to qumode-based devices by leveraging the Fock basis.
While both of these mappings can be realized by converting the QUBO to an
Ising Hamiltonian, the latter adds an additional degree of freedom in how
the problem is partitioned.  For example, a problem which would take 8 qubits
($2^8 = 256$ fock levels) could instead be mapped to qumodes with cutoffs of $2^{k_i}$
where $k_i$ is taken from the partition of 8 (excluding partitions which contain 1).
These partitions are shown in Table~\ref{tab:partitionsOf8}.
Choosing partition 4, we would have 2 qumodes each of $2^4 = 16$ fock levels while the full
represented state remains the same size as the original qubit-based problem. This flexibility
in terms of problem encoding can be exploited depending on various constraints (hardware limitations, 
grouping variables, etc).

\begin{table}[htb]
    \label{tab:partitionsOf8}
    \begin{tabular}{r|l}
        \hline \hline
        index & partition \\ 
        \hline
        1 & 2, 2, 2, 2 \\
        2 & 3, 3, 2  \\
        3 & 4, 2, 2  \\
        4 & 4, 4     \\
        5 & 5, 3     \\
        6 & 6, 2     \\
        7 & 8   \\
        \hline \hline
    \end{tabular}
    \caption{Partitions of 8 - excluding partitions containing 1.  Index 1 corresponds to the equivalent of 4 qubits, while Index 7 corresponds to a single mode of 8 levels.}
\end{table}

While we are limited in terms of problems for comparison, we
conjecture that the availability of qumodes will improve the efficacy
of VQA approaches since we gain ``new'' bosonic gates that can perform unique higher dimensional transformations.
Further, we can compare the various problem-aware ansatzes on these
various hardware configurations by leveraging various gates as unique
mixers. Note that the trainability and cost function landscape of the cost function for these bosonic devices may be very different \cite{zhang2024energy,dutta2025simulating}. One particularly interesting question is to quantify the expressivity and trainability of hybrid CV-DV circuits using tools from dynamical Lie algebra \cite{ragone2024lie,diaz2023showcasingbarrenplateautheory}.

Looking ahead to more scalable commercially-available qumode-based
hardware, having a survey and benchmark of the efficacy of different hardware
configurations for solving a particular class of optimization problem could be very helpful.

\subsubsection{CV Optimization via Quantum Dynamics}

One recent interesting optimization algorithm is the quantum Hamiltonian descent (QHD) formalism \cite{leng2023quantumhamiltoniandescent}. The idea is to map the optimization dynamics trajectory as a time-dependent quantum dynamics evolution governed by a Hamiltonian of many coupled anharmonic modes. 

Consider the following time-dependent effective Hamiltonian between $N$ bosonic modes
    \begin{align}
        \hat{H}(t) = \sum_{k=1}^N \frac{\hat{p}^2_k}{2 \mu(t)} + \sum_{j,k = 0}^{N} V_{jk} \hat{x}_j \hat{x}_k + \sum_{j,k,l,m = 0}^{N} W_{jklm} \hat{x}_j \hat{x}_k \hat{x}_l \hat{x}_m  ,
    \end{align}
    where $\mu(t) = 1+ t^2$ is an effective time-dependent mass (the time-dependency can change). Given an initial state of the parameter arrangement encoded in the state $\ket{\psi_0}$, the QHD algorithm claims that the time-dependent evolution of $\ket{\psi_0}$ for some time $t$ will converge to a good local minimum (or even global minimum) of the parameters, encoded in the final quantum state
    \begin{align}
        \ket{\psi(t)} = \mathcal{T} e^{-i \int_{0}^t \hat{H}(\tau) d\tau} \ket{\psi_0},
    \end{align}
    where $\mathcal{T}$ is the time-ordered product.

The main problem is then to figure out simple ways to decompose this time-dependent Hamiltonian simulation using hybrid CV-DV quantum signal processing on oscillator-qubit devices, for example, by using instructions from the phase space or Fock space ISAs. The recent progress on time-dependent Hamiltonian simulation algorithms \cite{low2019hamiltoniansimulationinteractionpicture,berry2019timedepedentHS,bosse2025efficient,sharma2024hamiltoniansimulationinteractionpicture,fang2025time,An2022timedependent,fang2025highordermagnusexpansionhamiltonian} can shed some light on solving this problem on CV-DV quantum hardware.

\subsection{Factoring}
\label{ssec:factoring}

Shor's algorithm achieves an exponential speedup for integer factoring by reducing the problem to finding the period \( r \) of the modular function \( f(x) = a^x \bmod N \), where \( a \) is chosen such that \(\gcd(a, N) = 1\). The quantum Fourier transform (QFT) can be used to find the period \( r \), which then enables efficient recovery of the prime factors of \( N \). 

In the CV-DV scheme introduced in \cite{cvdv-shors}, the period-finding stage is implemented using just three CV qumodes and a single qubit, regardless of the size of \( N \). This is made possible by exploiting the relationship between the position and momentum operators, which are Fourier transforms of each other, eliminating the need for an explicit QFT. All gates are applied in the position basis, and measuring the first qumode in the momentum basis at the end effectively obtains the one classical output of the Fourier transform. Note that if one instead wants the entire quantum output, then the QFT is still required explicitly, though can be achieved using oscillator free-evolution \cite{Liu_2025}. Fortunately, Shor's algorithm only depends on this classical output.

As mentioned above, Shor's algorithm in the CV-DV setting requires three qumodes. Two of the qumodes are initialized in GKP state \cite{GKP2001}, and the third qumode in the vacuum state. A GKP state $\ket{GKP}$ can be expressed as a uniform superposition of position eigenstates $\ket{GKP} \propto \sum_{x \in \mathbb{Z}} \ket{x}$. Therefore the initial state of the CV-DV system of three qumodes and one qubit is:
\begin{itemize}
    \item Qumode 1 in GKP State $\ket{GKP_{1}}  \propto \sum_{x \in \mathbb{Z}} \ket{x}$,
    \item Qumode 2 in GKP State $\ket{GKP_{2}}  \propto \sum_{y \in \mathbb{Z}} \ket{y}$,
    \item Qumode 3 in Vacuum State $\ket{0_{Osc}}$,
    \item Qubit in state $\ket{0}$.
\end{itemize}
There are three key steps in the algorithm as summarized below:
\begin{itemize}
    \item Applying a gate $M_N$ to Qumode 2. This is a multiplication gate that multiples the position eigenstate $y$ by $N$ to transform it to $yN$. This gate can be implemented using a single mode squeezing gate with squeezing parameter $\log N$.

    This gate changes the state of the system from 
    \[
    |\psi_1\rangle \ \propto \ \left( \sum_{x \in \mathbb{Z}} |x\rangle \right)
    \otimes
    \left( \sum_{y \in \mathbb{Z}} |y\rangle \right)
    \otimes
    |0_{\mathrm{Osc}}\rangle
    \otimes
    |0\rangle ,
    \] 
    to the state
    \[
    |\psi_2\rangle \ \propto \ \left( \sum_{x \in \mathbb{Z}} |x\rangle \right)
    \otimes
    \left( \sum_{y \in \mathbb{Z}} |y N\rangle \right)
    \otimes
    |0_{\mathrm{Osc}}\rangle
    \otimes
    |0\rangle .
    \]

    \item Applying the modular exponentiation unitary $U_{a,N}$ to Qumode 2. This unitary essentially adds the function $a^x \bmod{N}$ to the position eigenstates of Qumode 2. The exact decomposition of this unitary is described in \cite{cvdv-shors}. The state of the system is now
    \[
    |\psi_3\rangle \ \propto \ 
    \left(\sum_{x \in \mathbb{Z}} \ \sum_{y \in \mathbb{Z}} 
    \left| x \right\rangle \ \otimes \ \left| y N \ + \ \left( a^{x} \bmod N \right) \right\rangle\right)
    \otimes
    |0_{\mathrm{Osc}}\rangle
    \otimes
    |0\rangle .
    \]
    Note that, this entangles Qumode 1 and 2.
    \item Finally, we trace out Qumode 3 and the Qubit and perform a measurement on Qumode 2. The measurement of Qumode 2 will lead to an outcome \textit{k}. This means that Qumode 1 will collapse to states that obey the following property 
    \[
    y N + a^{x} \bmod N = k 
    \implies
    a^x \equiv k \pmod{N} .
    \]
    Therefore the state of Qumode 1 after measurement will be $\sum_{x \in \mathbb{Z}}|x\rangle_{a^x \equiv k \ \pmod{N}}$. Measuring this state in the momentum basis will directly provide the period of the function $a^x \bmod{N}$.
    
\end{itemize}

With this approach, the algorithm attains an \(\mathcal{O}(n^2)\) gate complexity for factoring an \( n \)-bit number, and \cite{cvdv-shors} specifies parameter ranges for \( R \), \( m \), and \( \Delta \). By contrast, a DV-only implementation of Shor's algorithm requires \(\mathcal{O}(n^{2} \log n)\) gates and uses $2n+2$ qubits to factor an $n-$bit number $N$~\cite{shors-resource}.

We note that more general Fourier transforms or character transforms \cite{unification2025bastidas} over groups \cite{rudin2017fourier}, especially continuous groups such as $SU(N)$, have the potential to be realized on CV systems more easily than mapping to DV systems.

\subsection{Quantum data compression and communication}
\label{ssec:compress-comm}

\begin{figure}[h]
    \centering
    \includegraphics[width=\linewidth]{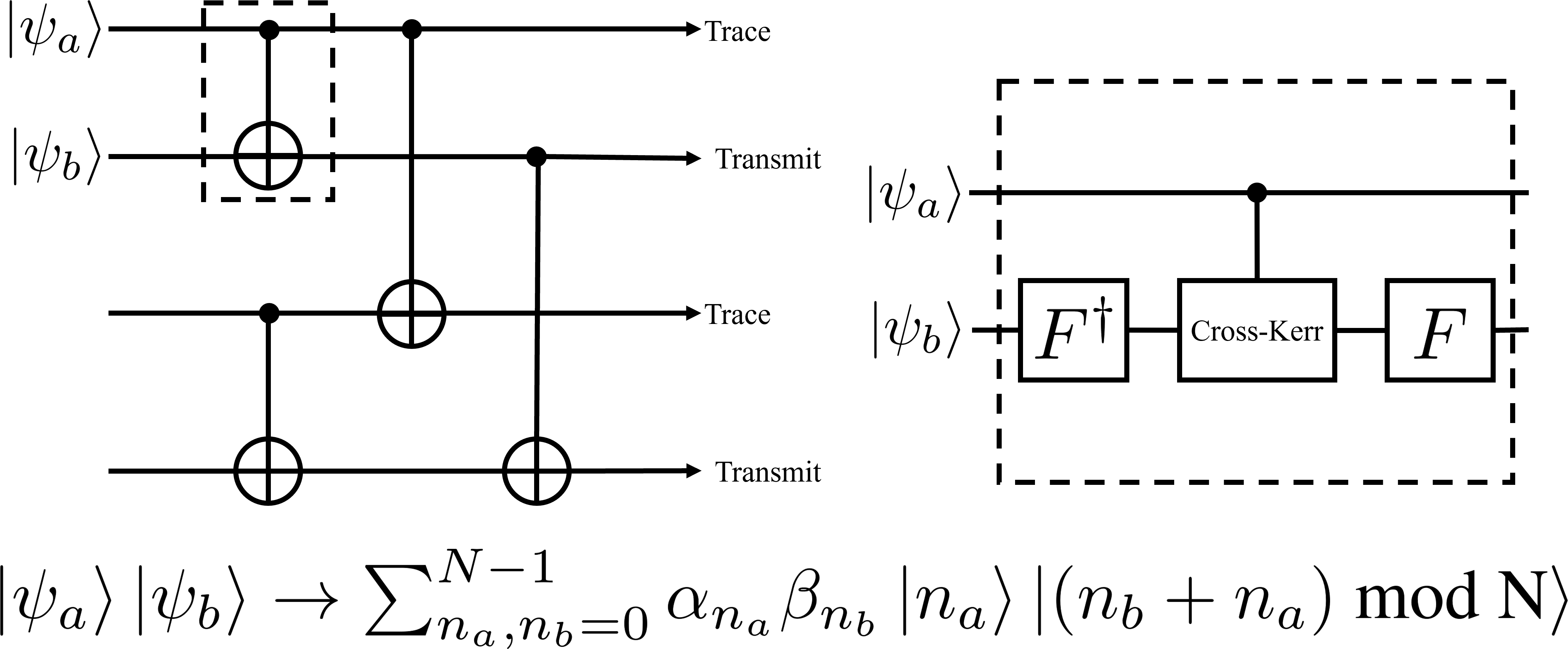}
    \caption{The polar code lattice and the direct translation of the modular addition operation to a quantum modular addition operator constructed by the inverse DFT on Fock levels, Cross-Kerr non-lineariy, and DFT.}
    \label{fig:polar_code}
\end{figure}

Any classical communication and memory will only be efficient if they leverage information-theoretic tools such as data compression~\cite{cover_elements_2001}. This is equally true for quantum systems. As most quantum communication channels are based on optics or photons, this motivates exploring compression protocols for CV quantum systems. 

In classical information theory, compression is usually framed in terms of simple bits, whereas the quantum CV setting naturally involves multi-level systems in superposition. This higher-dimensional structure poses significant new challenges.

For qubits, the foundational result is Schumacher compression, the quantum analogue of Shannon’s source coding theorem~\cite{schumacher_quantum_1995}. Schumacher’s protocol shows that quantum states emitted from a source can be compressed by encoding them into the typical subspace of the source. Our aim is to extend these ideas to qumodes. Fundamentally, there are two possible approaches: (i) convert qumodes into qubits and then apply Schumacher compression, or (ii) compress the oscillators directly. The first option introduces additional circuit overhead, noise, and resource demands. In contrast, oscillators already provide a discrete spectrum when truncated, and compression only requires a finite-dimensional Hilbert space. This makes direct compression of qumodes feasible.

As a concrete example, we focus on polar codes which have already been implemented for qubit systems~\cite{weinberg_quantum_2024}. Polar codes generalize beyond binary alphabets and extend naturally to $d$-ary systems, making them well suited for truncated oscillator encodings where each mode is restricted to a finite $d$-dimensional subspace. By translating the $d$-ary classical construction into quantum circuits, we obtain a Schumacher-style protocol for qumodes. The central challenges are not in the coding theory itself but in implementing the required quantum gates within CV hardware, illustrated in Fig.~\ref{fig:polar_code}. For example, modular addition is essential in polar coding but not a native oscillator operation under Fock basis.

To realize this, we exploit the modular structure of the phase states~\cite{royer_phase_1996},  the phase accumulated by an oscillator under free evolution is defined only modulo \(2\pi\). Thus, modular addition between qumodes in the Fock basis can be implemented by first applying a basis transformation into the phase basis, performing a conditional phase shift (e.g. using a cross-Kerr nonlinearity), and then undoing the basis transformation. It turns out that the discrete Fourier transform (DFT) maps Fock states to phase states. Since the DFT is a unitary operator, it can in principle be implemented using any universal gate set, even if it is not native to a given CV platforms. Finally, decoding a polar code is an iterative but closed-form algorithm, which admits translation into a sequence of unitary gates. This makes polar-code-based compression of qumodes not only theoretically sound but also realizable in principle within CV architectures.

Another key resource for quantum communication is entanglement. Entangled CV modes coupled to qubits can effectively serve as a source of Bell-pair generation. Ref.~\cite{PhysRevA.111.052610} proposes a non-deterministic scheme for generating Bell pairs using linear optics and measurement. More recently, deterministic state-transfer protocols have been shown to enable deterministic Bell-pair generation directly from entangled oscillators~\cite{Liu_2025}. These methods can also be used to perform entanglement swapping between CV and DV systems~\cite{PhysRevLett.114.100501, guccione2020connecting}.

\subsection{Ordinary differential equation solver}
\label{ssec:diffq}

Differential equations, such as the Navier–Stokes equations, are central to many areas of science and engineering. Despite the continuous nature of the variables and solutions in differential equations, classical differential equation solvers often relying on discretization. Hybrid CV-DV quantum processors, however, offer the possibility to solve these equations by mapping continuous variables and integrals directly onto CV registers, avoiding discretization~\cite{Jin_2024}.

A simple linear ordinary differential equation (ODE) takes the form
\begin{align}
    \frac{\dd u(t)}{\dd t} = -A(t)u(t) + b(t), \quad u(0) = u_0 ,  \label{eq:lchs-ODE}
\end{align}
where $t \in [0, T]$, $A(t) \in \Cbb^{N \times N}$, and $u(t), b(t) \in \Cbb^{N}$. The solution of the ODE~\eqref{eq:lchs-ODE} has the explicit form
{\footnotesize
\begin{align}
    u(T) &= \Tcal e^{- \int^T_0 A(s) \dd s}u_0 + \int^T_0 \Tcal e^{-\int^T_s A(s')\dd s'}b(s) \dd s ,  \label{eq:exp-sol}
\end{align}
}%
where $\Tcal$ is the time-ordering operator. By combining finite-difference schemes with linearization techniques, such as Carleman linearization~\cite{liu2021carleman}, a wide range of nonlinear partial differential equations can be transformed into a linear ODE system similar to the form described above.

For problem sizes relevant to real-world applications, classical solvers already operate at extreme scales. For example, the isotropic turbulence dataset in the Johns Hopkins Turbulence Database uses a grid of $32,768^3$ points~\cite{Li2008JHTDB}. By providing efficient access to the matrices and vectors in differential equations using techniques like block encoding~\cite{pocrnic2025constant}, quantum differential equation solvers have been developed for standard DV quantum computers. One promising example is the Linear Combination of Hamiltonian simulation (LCHS) algorithm~\cite{PhysRevLett.131.150603,an2023quantum}, which achieves optimal state-preparation cost and near-optimal matrix-query complexity across all relevant parameters. 

In LCHS, the coefficient matrix $A(t)$ in \eqref{eq:lchs-ODE} is first decomposed into $A(t) = L(t) + iH(t)$ via $L(t) = \left[A(t) + A^\dagger(t)\right]/2, H(t) = \left[A(t) - A^\dagger(t)\right]/2i$. It is assumed that $L(t) \succeq 0$. Otherwise, we can shift $u(t) = e^{ct}v(t)$ and use $v(t)$ for the ODE system so that $L(t) + cI \succeq 0$, where $-c$ is the minimum of the smallest eigenvalues of $L(t)$. The LCHS algorithm rewrites the two terms in the solution \eqref{eq:exp-sol} as integrals over parameterized Hamiltonian simulations
{\footnotesize
\begin{align}
    &\Tcal e^{- \int^t_0 A(s) \dd s} = \int_{\Rbb}  g(k) \left[\Tcal e^{-i \int^t_0 (kL(s)+H(s))\dd s} \right] \dd k \label{eq:LCHS-sol-homo} \\
    &\int^t_0 \Tcal e^{-\int^t_s A(s')\dd s'}b(s) \dd s \nonumber \\
    &\quad = \int^t_0 \int_\Rbb g(k) \left[ \Tcal e^{-i \int^t_s (kL(s')+H(s'))\dd s'} \right]  b(s) \dd k \dd s   \label{eq:LCHS-sol-inhomo} , 
\end{align}
}%
where $g(k):=\frac{f(k)}{1-ik}$ for some kernel function $f(k)$ and $t \in [0,T]$. The choice of $f(k)$ that leads to near-optimal complexity dependency on all parameters is 
$
    f(z) = \frac{1}{(2\pi e^{-2^{\beta}}) e^{(1+iz)^\beta}},
$
where $\beta \in (0,1)$ is a user-defined constant~\cite{an2023quantum}.
Note that the terms in brackets in Eq.~\eqref{eq:LCHS-sol-homo} and~\eqref{eq:LCHS-sol-inhomo} are unitary. 

The standard DV LCHS algorithm further approximates the integrals over $k$ in Eqs.~\eqref{eq:LCHS-sol-homo} and~\eqref{eq:LCHS-sol-inhomo} by discretizing and truncating them into a finite sum of unitaries. This discretization incurs substantial quantum-resource overhead in circuit implementation, making it costly for both near-term devices and early fault-tolerant quantum machines. In particular, 
the number of terms for discretizing the integration, denoted by $M$, scales as
\begin{align*}
        M \in \Ocal\left( T \max_t \|L(t)\|\left(\log \frac{1}{\epsilon}\right)^{1+1/\beta} \right) , 
\end{align*}
for $\epsilon$ being the total error incurred by LCHS. The corresponding circuit requires $\lceil\log_2(M)\rceil$ ancillary qubits, and more problematically, $\lceil\log_2(M)\rceil$-controlled multiple-qubit unitary operators for Hamiltonian simulations. In this regime, for a time-independent ODE, the value of $M$ can reach the order of $10^6$ when $\epsilon$ is around $10^{-5}$, $\beta=0.8$, $T=1000$, and $\|L\|=1$. More exact scalings of LCHS are provided in~\cite{pocrnic2025constant}. 

By encoding the distribution $g(k)$ as wave functions on CV systems, the integration in Eqs.~\eqref{eq:LCHS-sol-homo} and \eqref{eq:LCHS-sol-inhomo} can be performed by measurement on the CV register. Ref. \cite{bell2025codesigningeigensingularvaluetransformation} provides an example of a CV-based LCHS approach for ground-state preparation in spin models, and the same underlying idea can be applied to ODE solvers as well.

\subsection{Sensing}
\label{ssec:sensing}
Quantum sensing \cite{RevModPhys.89.035002} utilizes quantum resources such as quantized energy levels, coherent superpositions, and entanglement to measure physical quantities, achieving parameter sensitivity beyond the capabilities of classical protocols \cite{hb3c-dk28,ligo, doi:10.1021/nl402552m, rugar2004, RevModPhys.86.1391}. Qubit-based sensing can surpass the limits of classical sensors by exploiting quantum entanglement and coherence \cite{doi:10.1126/science.1104149}.

Hybrid CV-DV quantum systems offer a potentially more powerful platform for quantum sensing than purely qubit-based approaches. One advantage is that the infinite-dimensional Hilbert space of a single CV mode allows it to act as a highly efficient transducer, encoding significantly more information about a signal without the need for a large entangled qubit register. A second advantage is that information encoded in the CV mode can then be effectively manipulated via coupling to a DV register \cite{liu2024hybrid} and then readout by performing standard measurement on the DV qubits. There is also work discussing how CV methods can be used to model the noise affecting DV quantum sensing circuits~\cite{jebraeilli2025stqs}.

There has been a growing interest in exploring various bosonic-mode-enhanced quantum sensing protocols, such as those involving cat-states or superpositions of Fock states. Ref. \cite{SinananSingh2024singleshotquantum} introduced quantum signal processing interferometry (QSPI), which integrates qubits and bosonic modes via block-encoded hybrid gates to perform binary decision tasks, such as weak-signal detection, using single-qubit readout, enabled by computation performing on the hybrid CV-DV system. QSPI achieves Heisenberg-limited (HL) scaling across the full parameter regime of interest for binary decision problems more resource efficiently than cat-state-based sensing.
Ref.~\cite{PRXQuantum.6.010304} demonstrates an efficient phase estimation protocol on a circuit-QED platform that uses superpositions of coherent states and performs two state-transfer operations on a single bosonic mode.
Another work~\cite{ng3t-p1n1} uses cat states to estimate frequency shifts, achieving HL sensitivity by using phase-space rotations and displacements to map the frequency shift onto a measurable phase difference between the cat-state components, which is then read out by a qubit.

\subsection{Fault tolerance}
\label{ssec:ft}

The theory of quantum error correction (QEC) and fault tolerance is essential to turning controlled quantum operations into reliable quantum computation. QEC protocols that encode DV quantum information into larger discrete or continuous Hilbert space \cite{Chuang1997,Gottesman:2000di,Binomial2016, Hu2019,Cat2013, Ofek2016,Liu:2024mbr,Aoki:2009dgq,albert_2025} have been proposed, 
with the latter approach known as bosonic QEC~\cite{PRXQuantum.2.020101}. Hybrid CV-DV systems offer easier methods for logical state preparation and logical gate synthesis, potentially combining the advantages of DV-to-DV and DV-to-CV encodings~\cite{ruiz2025unfoldeddistillationlowcostmagic}. A key challenge is scaling these techniques from small demonstrations to large, utility-scale implementations involving multiple qumodes and qubits
~\cite{PRXQuantum.3.010335,xu2025lettingtigercagebosonic}. 

A much less explored direction is to encode robust (not necessarily ``logical" in the same sense as logical qubits) CV degrees of freedom into more CV modes, i.e., CV-to-CV encoding. 
Refs.~\cite{PhysRevLett.80.4084, PhysRevLett.80.4088} explored this possibility where the idea of repetition codes was used as an encoding scheme that distributes the logical variable redundantly across multiple wavepackets. While Lloyd and Slotine~\cite{PhysRevLett.80.4088} originally constructed a nine-wavepacket code capable of correcting arbitrary single wavepacket displacement errors, Braunstein subsequently improved it to a five-wavepacket code~\cite{PhysRevLett.80.4084}. This construction closely parallels the discrete $[[5,1,3]]$ perfect QEC code~\cite{Calderbank1996, Steane:1995vv,Laflamme1996}, encoding one mode into five and correcting arbitrary displacement errors on any one mode.

These early successes have spured more recent advancements. It was realized that errors on CV modes produced by Gaussian channels (for example, random displacement channels on each qumode) cannot be corrected with only Gaussian resources. In other words, any encoding scheme that does not consume non-Gaussian states or does not use non-Gaussian gates will not be able to correct Gaussian errors \cite{PhysRevA.99.032344}. It was further shown that encoding schemes that do not contain non-Gaussian resources cannot exhibit a fault-tolerant threshold in CV systems~\cite{Hanggli_2022}, in sharp contrast to standard DV quantum error correction. Indeed, using non-Gaussian resources such as GKP states, it is possible to suppress errors in Gaussian channels \cite{noh2020encoding}, despite the requirement to prepare a fresh GKP state for every cycle of error detection. Further questions remain open regarding which sets of logical CV operators can be implemented transversally and which cannot.
Beyond introducing space-redundancy for robust CV encoding, ideas from control theory \cite{brogan1985modern} may be leveraged for improved robustness of hybrid CV-CV computing without significant spatial overhead.

Although achieving a truly logical, infinite-dimensional CV-to-CV encoding may be an unrealistic goal, practical utility does not necessarily require fully fault-tolerant CV quantum computation.  Real-world application, such as those described in Secs. \ref{sec:natural-science} and \ref{sec:other-computing}, will be executed with finite time, finite precision, and finite energy. Therefore, if a CV-to-CV encoding is sufficiently robust to provide clear utility, the objective is achieved. In particular, a finite-dimensional encoding with simple logical ``quadrature" operators that approximately satisfy the bosonic commutation relations will be useful.

%%%%%%%%%%%%%%%%%%%%%%%%%%%%%%%%%%%%%%%%%%%%%%%%%%%%%%%%
\section{Algorithms, Software, and Techniques}
\label{sec:alg-soft-tech}

In this section we discuss some tools and techniques unique to hybrid CV-DV quantum computing that are useful for addressing utility problems. We describe two basic quantum-algorithmic primitives on hybrid CV-DV processors in Sec.~\ref{ssec:alg}. This is followed by a brief overview of the quantum software landscape in Sec.~\ref{ssec:software} and classical simulation of hybrid CV-DV quantum computing in Sec.~\ref{ssec:classical-simulation}.

\subsection{Hybrid CV-DV algorithms: from quantum control to algorithms and back}
\label{ssec:alg}

The distinct characteristics of CV systems make algorithm design for hybrid CV–DV processors nontrivial. Nonetheless, the guiding principles of the leading DV algorithms can often be extended to CV–DV settings. Recent works exemplify this trend by generalizing Trotter and product formulas \cite{kang2023leveraging}, linear combinations of unitaries \cite{Chakraborty2024implementingany,bell2025codesigningeigensingularvaluetransformation}, and quantum signal processing \cite{SinananSingh2024singleshotquantum,Liu_2025,singh_towards_2025} to hybrid architectures. For an overview, we refer the reader to a recent tutorial \cite{liu2024hybrid} and to applications of CV–DV processors in quantum simulation \cite{crane2024hybridoscillatorqubitquantumprocessors}. It is also worth noting that early quantum control techniques, such as composite pulses and NMR methods, inspired a wide range of algorithmic developments that now are returning to improve programmable control of CV–DV systems with rigorous performance guarantees. In this hybrid context, the traditional boundaries between control and algorithms are beginning to blur.

In this section, we present two additional new algorithm constructions that are enabled by hybrid CV-DV quantum computing. Sec.~\ref{sssec:rotor-qpe} describes a phase estimation algorithms constructed by hybridizing qubits and CV quantum rotors. Sec.~\ref{sssec:cv-lchs} highlights a CV linear combination of Hamiltonian simulation (LCHS) algorithm with applications to solving differential equations.

\subsubsection{Quantum Phase Estimation with Rotors}
\label{sssec:rotor-qpe}

Even rarer in quantum computing than oscillators are rotors, which are
modeled as particles constrained to move along a ring. Like
oscillators, rotors have position and momentum degrees of freedom, here corresponding to angular position and angular
momentum. 
We denote the angular position operator as \( \hat{\varphi} \) and the
angular momentum operator as \( \hat{l} \). Unlike oscillators and qudits,
rotors have the property that the position space is not isomorphic to the
momentum space. 
Because the base space of the rotor is \( \mathrm{U}(1) \), it is naturally suited for simulating systems with periodic structure or 
underlying \( \mathrm{U}(1) \) symmetry, such as certain gauge theories.
Moreover, since the Pontryagin dual of \( \mathrm{U}(1) \) is \( \Z \), and \( \Z \)  
is not isomorphic to \( \mathrm{U}(1) \), some standard quantum algorithmic primitives
cannot exist in their usual form. As a result, algorithms like the QFT must be adapted
using combinations of position and momentum measurements.

Quantum Phase Estimation (QPE) is an approximate algorithm for recovering the phase
applied on an eigenvector of a unitary operator \cite{Nielsen_Chuang_2010}. More
precisely, for some unitary operator $\hat{U}$ and known eigenvector
$\ket{\psi}$, the QPE algorithm computes \( \theta \), where \( \hat{U}
\ket{\psi} = e^{i \theta} \ket{\psi} \).
Notably, for the execution of QPE, there is a requirement of
performing a Quantum Fourier Transform on the ancilla register qubit. The final
output is given by a measurement of the ancilla qubits, which returns the bitstring
representing \( \theta \) with at least \( \frac{4}{\pi^2} \) probability. 

We can perform this same QPE algorithm by replacing the ancilla register with a single quantum rotor (a CV system). Instead of an oracle
that computes controlled \( U^{2^j} \) operations, however, we need an oracle
that performs the following:
\begin{equation}
    \hat{O} = \sum\limits_{l \in \Z} \ket{l}\bra{l} \otimes U^{l},
    \label{eq:rotor-qpe-oracle}
\end{equation}
where $\ket{l}$ is a state that labels the quantized angular momentum of the rotor.

\begin{figure}[h]
    \centering
    \includegraphics[width=0.6\linewidth]{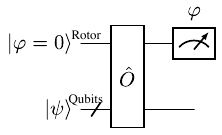}
    \caption{Quantum Phase Estimation algorithm using a single rotor with the
    oracle given in Equation \eqref{eq:rotor-qpe-oracle}. %\textcolor{blue}{Yuan: can we emphasize in the figure which register is a qubit, which is a rotor explicitly?}
    }
    \label{fig:rotor-qpe}
\end{figure}
For any eigenstate of the phase operator, \( \ket{\varphi = \theta_0} \), the
action of the oracle behaves as an angular displacement:
\begin{align*}
    \hat{O} (\ket{\varphi = \theta_0}
    \otimes \ket{\psi}) & = \sum\limits_{l \in \Z} e^{il\theta}
    \braket{l|\varphi = \theta_0}(\ket{l} \otimes \ket{\psi}) \\
    & = \frac{1}{\sqrt{2 \pi}} \sum\limits_{l \in \Z} e^{il(\theta + \theta_0)}
    (\ket{l} \otimes \ket{\psi}) \\
    & = \ket{\varphi = \theta + \theta_0} \otimes \ket{\psi}.
\end{align*}
This hybrid CV-DV quantum phase estimation algorithm has a few advantages. Notably, no QFT has to be performed because the angular momentum eigenstate $\ket{l}$ and the phase states $\ket{\varphi}$ are related by a Fourier transform themselves. Furthermore, the
resolution of the algorithm, i.e., the precision of the measured \( \theta \), is
not dependent on the number of rotors, unlike qubit QPE, which requires a larger
qubit register for increased precision. Here, precision comes from phase
measurements and oracle implementation.

Moreover, implementing the oracle does not require a countably infinite linear combination of operators. With access to arbitrary-precision phase measurements, it suffices to define the oracle on the subset of momentum states supporting the approximate initial state, introducing a tradeoff between shot complexity and algorithmic overhead.

\subsubsection{CV-DV Linear Combination of Hamiltonian Simulation}
\label{sssec:cv-lchs}

Inspired by the continuous LCU framework of~\cite{bell2025codesigningeigensingularvaluetransformation}, a hybrid CV-DV formulation of LCHS offers a promising avenue for reducing resource costs and, ideally, eliminating the truncation and discretization errors inherent in the integral approximations used in standard qubit-based LCHS, as in Eqs.~\eqref{eq:LCHS-sol-homo} and~\eqref{eq:LCHS-sol-inhomo}. 
Specifically, continuous LCU can replace the $\lceil\log_2(M)\rceil$ ancillary qubits required in DV LCU circuits with $\Ocal(1)$ ancillary oscillators, together with CV state preparation and CV measurements, while leaving the Hamiltonian-evolution unitaries in the qubit space. 

To illustrate the process, let us consider a time-independent homogeneous ODE system, where $A(t):=A$ and $b(t) := 0$ for all $t$. From Eqs.~\eqref{eq:exp-sol} and~\eqref{eq:LCHS-sol-homo}, the complementary solution becomes
\begin{align}
    u(t) &= \left[\int_{\Rbb}  g(k)  e^{-i t (kL+H)}  \dd k \right] u_0 \label{eq:sol-homo-lchs} , 
\end{align}
for $t \in [0,T]$. Using a single oscillator for the ancilla in continuous LCU, the qubits that synthesize the unitary $e^{-i t (kL+H)}$ need to couple with this ancillary oscillator, leading to the joint evolution
\begin{align}
    e^{-it ( L \otimes \hat{p} + H \otimes I)}  ,
    \label{eq:joint-evo}
\end{align}
where $\hat{p}$ is the momentum operator and $I$ is the identity operator. That is to say, the qumode remains untouched when $H$ acts on the qubit system. 

Before forming the circuit for the joint evolution, the function $g(k)$ has to be obtained. To do this, we need to prepare and post-select two appropriate CV states. Let $\ket{\psi}$ be the preparation state, the state prepared in the oscillator before the joint evolution, and $\ket{\phi}$ be the projection state, the state post-selected at the end. The effective operation on the qumode register after the projection of the DV register to $\ket{\phi}$ is~\cite{bell2025codesigningeigensingularvaluetransformation}
\begin{align*}
     \bra{\phi}  e^{-it ( L \otimes \hat{p} + H \otimes I)} \ket{\psi} = \int_{\mathbb{R}} \dd p \, \phi^*(p) \psi(p)    e^{-it ( pL + H)}.
\end{align*}
Comparing to Eq.~\eqref{eq:sol-homo-lchs}, this gives $g(p)= \phi^*(p) \psi(p)$. Namely, we just need to engineer the preparation and post-selection states to obtain an appropriate $g(p)$ that satisfies LCHS's requirements. We can fix the post-selection state $\ket{\phi}$ to a squeezed vacuum state $\phi_r(p):=S(r,0)\phi_0(p)\propto e^{-p^2/e^{2r}}$, and this gives the corresponding initial  state to be prepared 
\begin{align}
    \psi_r(p):=g(p)e^{p^2/e^{2r}}.
\end{align}
We now consider two special cases of $\psi_r(p)$ preparation and their implications.
\begin{itemize}
    \item $\mathbf{r=0:}$ In this case $\phi_r(p)$ is the vacuum state. However, for the vacuum state, our preparation state becomes
    \begin{align}
        \psi_0(p) := g(p)e^{p^2} , 
    \end{align}
    which is not a physically realizable state, as $\int_\mathbb{R}|\psi_0(p)|^2dp$ is not finite and diverges. 
    
    This gives an additional condition for $g(p)$ for CV-DV LCHS:
\begin{align*}
    \exists\ r\in\mathbb{R}_+\ \text{s.t}\ \int_\mathbb{R}\mathrm{d} p|g(p)|^2e^{2p^2/e^{2r}}<\infty .
\end{align*}
    \item $\mathbf{r\to\infty:}$ This leads to $\phi_r(p)$ being the infinitely squeezed vacuum state. Then the preparation state is simply the kernel itself:
    \begin{align}
        \psi_\infty(p) := \lim_{r\to\infty} \psi_r(p)=g(p) .
    \end{align}
\end{itemize}
If necessary, the desired state $\ket{\psi}$ can be approximated via a finite Fock expansion $\psi(p) \approx \sum_{n = 0}^{N_{max}} C_n \braket{p|n},
$
where coefficients $\{C_n\}$ are defined as $C_n = \int_{\mathbb{R}} \dd p \, \phi^*_n(p) \psi(p).$
To achieve an overall approximation error $\epsilon$ in representing $\psi(p)$, it can be shown that choosing the Fock cutoff
\begin{align}
    N_{max} = O\Bigg(\log^2\Big(\frac{1}{\epsilon-O(e^{-2r})}\Big)\Bigg) , 
\end{align}
is sufficient for a finite squeezing parameter $r=\Omega(\log\frac{1}{\epsilon})$.

Synthesizing the oracle in Eq.~\eqref{eq:joint-evo} for matrices $L$ and $H$ is the key to make the entire algorithm efficient, with basis-gate implementations that are problem-specific and highly dependent on the computational basis chosen. Many previous works in the DV regime tackle this problem by decomposing coefficient matrices into linear combinations of Pauli matrices, for example, hyperbolic partial differential equations~\cite{PhysRevResearch.6.033246} and advection-diffusion equations~\cite{alipanah2025quantum}. Leveraging these design principles and following the circuit-synthesis techniques of~\cite{bell2025codesigningeigensingularvaluetransformation}, the evolution of decomposed Hamiltonian coupled to the oscillator can be implemented as a CV-DV circuit.

\subsection{Software development kit (SDK) for hybrid CV-DV computing}
\label{ssec:software}

A large software ecosystem exists for DV circuits (Qiskit~\cite{QiskitAer}, Cirq~\cite{Cirq}, Pennylane~\cite{bergholm2022pennylane}, etc.), enabling transpilation to hardware, classical simulation of small problem instances, and quantum resource estimation of large circuits, which is essential for determining the feasibility of quantum applications on real hardware. For CV systems, some limited options exist (for insatance Pennylane/StrawberryFields~\cite{strawberryfields}, Bosonic Qiskit~\cite{stavenger2022c2qa}).

To the best of our knowledge, only Bosonic Qiskit supports hybrid CV-DV instructions, but because it wraps Qiskit, it lacks native support for qumodes. One thrust of our work is developing a new package enabling similar capabilities already enjoyed by the DV ecosystem: abstractly defining hybrid CV-DV circuits, dispatching circuits to many different backends, and performing quantum resource estimation of large circuits.

Our early prototype~\footnote{is available at \url{https://www.github.com/pnnl/hybridlane} under a BSD 3-clause open-source license} extends Pennylane, as it was the only mainstream framework flexible enough to allow native qumode objects, and has several features:

\begin{enumerate}
    \item Gate definitions are decoupled from the matrices that represent them during simulation, enabling description and decomposition of circuits with large numbers of qubits and qumodes.

    \item Circuits are validated by statically analyzing the structure and inferring qumodes and qubits from the operations and measurements. Qubit/homodyne/Fock measurements are additionally inferred through static analysis.

    \item It is compatible with existing Pennylane gates and operators. Furthermore, hybrid CV-DV devices may be defined through the usual Pennylane ``Device'' interface, and we provide an example device that classically simulates these circuits using Bosonic Qiskit as a backend.

    \item An intermediate representation based on OpenQASM is also planned, with modifications to support typing qubits/qumodes and performing homodyne or Fock measurements.
\end{enumerate}

As development continues, we hope others will find it useful to integrate simulators and/or real hardware devices.

\subsection{Classical simulation of CV-DV circuits}
\label{ssec:classical-simulation}
General-purpose DV simulators such as Qiskit-Aer~\cite{QiskitAer}, Cirq~\cite{Cirq}, and QuEST~\cite{quest} have
pushed state-vector simulation to the scale of dozens of qubits, leveraging GPU acceleration~\cite{cuquantum} and
distributed HPC systems with strong-scaling
results demonstrated across multiple platforms~\cite{superqubit, c3pq, atlas}.
These successes highlight the maturity of DV simulation, while also
setting a baseline against which hybrid extensions must be measured.

The challenge becomes apparent when considering the scaling of hybrid
CV--DV systems. The statevector dimension grows as the product of
contributions from qubits and qumodes. For $n$ qubits, the Hilbert
space dimension is $2^n$, while for $m$ qumodes with cutoffs
$f_1, f_2, \dots, f_m$, it is
$f_1 \times f_2 \times \dots \times f_m$, or $f^m$ if each qumode has
the same cutoff $f$. Thus, the combined Hilbert space dimension is
$2^n \cdot f^m$, and when noise is included, the corresponding density-matrix dimension becomes
$4^n \cdot f^{2m}$. This exponential growth makes naive extensions of
DV tools to hybrid CV--DV settings impractical.

For CV-only systems, tensor network methods such as MPS~\cite{9605325,michelsen2025functionalmatrixproductstate}
provide partial relief by trading accuracy for efficiency, but
equivalent strategies for hybrid systems remain largely
unexplored. Reference~\cite{michelsen2025functionalmatrixproductstate} proposes an approach to represent the state vector as a tensor-train decomposition, saving on memory requirements and hence being able to store more qumodes compared to classical approaches. Furthermore, this work outlines that most of the gates used for this representation, e.g. displacement, squeezing, and beamsplitter, exhibit sparsity patterns which reduce both their corresponding memory requirements and computational complexity. One drawback of the tensor representation is the fact that the state vector must always be preserved in the tensor train format even after applying the gates. Therefore, as part of this thrust we focus on developing an approximate approach built on top of tensor-train decompositions. We propose an automatic framework meant to schedule the gate computation andthe re-compression stages, minimizing communication and redundant computation. We will focus on creating a solution that can easily be targeted towards both multi-CPU and GPU systems.

To address bosonic modes directly, CV-focused simulators such as
Strawberry Fields, Piquasso, and Perceval support photonic circuit
design with Gaussian and Fock
backends~\cite{strawberryfields,piquasso,perceval,graphiq,mrmustard}.
These frameworks excel at modeling continuous-variable processes but
typically impose a global cutoff and lack integration with qubits,
making them unsuitable for hybrid workloads.

As seen in the previous section, hybrid approaches are beginning to 
appear, most prominently
Bosonic-Qiskit, which extends Qiskit with qumodes and CV
gates~\cite{stavenger2022c2qa}.  However, it embeds each cutoff $f$ into
the nearest power-of-two qubit representation, overshooting the
required dimension. For example, a qumode with $f=12$ is mapped to
$x=4$ qubits (since $2^4=16$), thereby increasing the simulation
cost. Moreover, because it inherits Aer as its backend, every qumode
operation must be transpiled into a qubit circuit---a costly step
exacerbated in parametric workloads such as VQAs. Compiler-level
efforts like Genesis~\cite{genesis} explore CV--DV-QASM generation, but
a scalable simulator designed natively for hybrid CV--DV circuits has
yet to emerge. 

Hence, one thrust of our work is developing a general-purpose hybrid
CV--DV simulator aimed at HPC-scale execution. It operates directly on
mixed state spaces with qubits and bosonic modes, each qumode
supporting an independent Fock cutoff to optimize memory usage and
avoid the global cutoff bottleneck of existing Fock-based
simulators. Our package is being designed to scale across multi-node
HPC systems via MPI and to support multiple execution backends for
high-performance evolution over large state spaces. To encourage
adoption, we plan to make it interoperable with frontends such as
PennyLane, Qiskit, and our early natively hybrid
prototype (Section~\ref{ssec:software}) by implementing standard 
device/backend
interfaces, while also providing a native interface for advanced users
who wish to leverage HPC-specific optimizations directly.

\section{Conclusion}
\label{sec:conclusion}

The hardware capabilities of hybrid CV--DV quantum computers continue to advance, either as a natural component of the CV system, such as the vibrational modes in trapped ion systems, or in their own right, such as resonant cavities coupled to a DV mode. In this work, we have outlined a set of problems and research directions that can directly benefit from such architectures.  We have demonstrated that some of the most scientifically compelling and practically relevant problems are inherently hybrid, and therefore stand to gain substantially from quantum computers whose native CV–DV structure mirrors that of the target models.  Our highlighted problems range from applied (photo-chemistry) to theoretical (quantum field theory) to optimization (MaxCut), reflecting a broad scientific scope of interest across physics, chemistry, and computer science. We have further presented several new CV-DV quantum algorithms, and emphasized the need for corresponding software development and classical simulation capabilities. We hope that these problems will serve as useful benchmark targets for the community as hybrid hardware and software stacks continue to mature.

\vspace{0.2in}

\stoptoc
\acknowledgments
This work is supported by the U.S. Department of Energy, Office of Science, Advanced Scientific Computing Research, under contract number DE-SC0025384. This research used resources of the National Energy Research Scientific Computing Center (NERSC), a U.S. Department of Energy Office of Science User Facility located at Lawrence Berkeley National Laboratory, operated under Contract No. DE-AC02-05CH11231.

\section*{Author contributions}
MA, BB, JB, ERD, RGJ, AK, AFK, KK, AL, YL, MT, and MZ wrote the natural science applications section. The section on computing relevant applications was composed by BB, DB, ERD, FG, RGJ, AL, YL, AM, CM, SM, COM, HZ, and MZ.  The algorithms, software, and techniques were contributed by BB, SC, ERD, JF, AK, AL, FM, COM, MT, DTP, TS, and MZ.  All authors contributed to editing the manuscript.  The project was organized and led by AFK and YL.

\appendix

\section{Conical Intersections}
\label{app:CI}
Diabatic Hamiltonians are widely used to model conical intersections (CIs). The most general is the multi-state linear vibronic coupling (LVC) or quadratic vibronic coupling (QVC) \cite{doi:10.1021/jp994174i, doi:https://doi.org/10.1002/9780470142813.ch2, MARTINEZ1997139} form:
\begin{align}
H &= T_{\rm nuc} + \sum_{\alpha=1}^{M}\!\Big[E_\alpha 
+ \mathbf{k}_\alpha^\top \mathbf{q} + \tfrac12 \mathbf{q}^\top \mathbf{K}_\alpha \mathbf{q}\Big]\,|\alpha\rangle\langle\alpha| \nonumber\\
&\quad+ \sum_{\alpha<\beta}\!\Big[\mathbf{\lambda}_{\alpha\beta}^\top \mathbf{q} 
+ \tfrac12 \mathbf{q}^\top \mathbf{\Lambda}_{\alpha\beta} \mathbf{q}\Big]\big(|\alpha\rangle\langle\beta|+{\rm h.c.}\big),
\end{align}
where $T_{\rm nuc}=\sum_i p_i^2/2m_i$ and $\mathbf{q}$ are mass-weighted normal coordinates (optionally including Duschinsky rotations). The diagonal blocks describe local PESs (LVC/QVC), while off-diagonal couplings generate the CI seam and branching plane.  

For minimal models, a two-state, two-mode LVC Hamiltonian in the branching-plane basis is often used:
\begin{align}
H &= T_{\rm nuc} + \tfrac12\!\left(P_g^2+P_h^2\right)\mathbf{1}
+ \tfrac12\!\left(\Omega_g^2 Q_g^2+\Omega_h^2 Q_h^2\right)\mathbf{1} \nonumber\\
&\quad + \big[\Delta + \kappa_g Q_g\big]\sigma_z
+ \lambda_h Q_h\,\sigma_x,
\end{align}
where $Q_g$ tunes the energy gap, $Q_h$ provides the nonadiabatic coupling, and $\sigma_{z,x}$ act on diabatic states.  

In open-system contexts, a spin–boson type Hamiltonian is used to include dissipative environments:
\begin{align}
H &= H_{\rm LVC} + \sum_k \omega_k b_k^\dagger b_k 
+ \sum_{\alpha,k} g_{\alpha k} |\alpha\rangle\langle\alpha|(b_k^\dagger+b_k),
\end{align}
which couples electronic states to harmonic baths representing protein or solvent fluctuations. CIs are characterized by branching-plane vectors in the adiabatic representation: $\mathbf{g}=\nabla_{\mathbf{R}}(E_2-E_1)$ and $\mathbf{h}=(E_2-E_1)\,\mathbf{d}_{12}$ with $\mathbf{d}_{12}=\langle\phi_1|\nabla_{\mathbf{R}}\phi_2\rangle$. Dynamics are tracked by population transfer $P_{\alpha\!\to\!\beta}(t)$, branching ratios, quantum yields, nonadiabatic coupling norms $\|\mathbf{d}_{\alpha\beta}\|$, geometric phase signatures, and spectroscopic observables such as ultrafast pump–probe or 2D electronic signals.

\resumetoc
\bibliography{ref,propref}

\end{document}